\documentclass[%
 aip,
 amsmath,amssymb,
 reprint,%
]{revtex4-1}
\pdfoutput=1
\usepackage{graphicx}% Include figure files
\usepackage{dcolumn}% Align table columns on decimal point
\usepackage{dsfont}
\usepackage{bm}% bold math
\usepackage[utf8]{inputenc}
\usepackage[T1]{fontenc}
\usepackage{mathptmx}
\usepackage{etoolbox}

\usepackage{xspace}
\usepackage{array}
\usepackage{stmaryrd}
\newcolumntype{K}[1]{>{\centering\let\newline\\\arraybackslash\hspace{0pt}}m{#1}}
\usepackage{makecell}
\usepackage{float}
\usepackage{soul}
\makeatletter
\def\@email#1#2{%
 \endgroup
 \patchcmd{\titleblock@produce}
  {\frontmatter@RRAPformat}
  {\frontmatter@RRAPformat{\produce@RRAP{*#1\href{mailto:#2}{#2}}}\frontmatter@RRAPformat}
  {}{}
}%

\usepackage[mathscr]{euscript}
\usepackage{color}
\usepackage{xcolor}
\usepackage{xspace}
\usepackage{amsmath}
\usepackage[bb=boondox]{mathalfa}
\usepackage{mathrsfs}
\usepackage{multirow}

\definecolor{darkgreen}{rgb}{0,0.5,0}

\newcommand{\uL}{\ensuremath{\,\mu\mathrm{L}}\xspace}

\DeclareMathAlphabet{\mymathbb}{U}{BOONDOX-ds}{m}{n}

\renewcommand{\deg}{\ensuremath{{}^{\circ}}\xspace}

\newcommand{\abs}[1]{|#1|}

\def\Cth{\ensuremath{{}^{13}\mathrm{C}}\xspace}
\def\Ctwo{\ensuremath{{}^{13}\mathrm{C}_2}\xspace}

\newcommand{\wJ}{\omega_{J}}
\newcommand{\Tone}{T_1}

\newcommand{\iu}{{i\mkern1mu}}

\newcommand\ket[1]{|{#1}\rangle}
\newcommand\bra[1]{\langle{#1}|}

\newcommand{\hf}{\tfrac{1}{2}}

\newcommand{\boldI}{\boldsymbol{I}}

\newcommand{\wSTnut}{\omega^{\rm ST}_\mathrm{nut}}

\newcommand{\Rsymm}[3]{\ensuremath{\mathrm{R}#1_{#2}^{#3}}\xspace}
\newcommand{\RNnnu}{\ensuremath{\mathrm{R}N_n^\nu}\xspace}

\newcommand{\Tzz}{$\mathrm{T_{00}}$\xspace}
\newcommand{\wS}{\omega_\Sigma}
\newcommand{\wD}{\omega_\Delta}
\renewcommand{\wJ}{\omega_J}
\newcommand{\wnut}{\omega_{\rm nut}}

\newcommand{\Hrf}{H_\mathrm{rf}}
\newcommand{\disopropanol}{isopropanol-$\emph{d}_8$\xspace}
\newcommand{\tauexc}{\tau_\mathrm{exc}}
\newcommand{\citeLLS}{\cite{%
pileio_long-lived_2020,%
carravetta_LFsinglet_2004,%
carravetta_HFsinglet_2004,%
sarkar_sequence_2007,%
pileio_storage_2010,%
tayler_singlet_2011,%
MHL-LLS-ARPC_2012,%
tayler_accessing_2013,%
devience_preparation_2013,%
devience_nuclear_2014,%
devience_probing_2016,%
devience_homonuclear_2021,%
Levitt-LLSS_2019,%
zhang_long-_2014,%
zhang_singlet_2015,%
stevanato_nuclear_2015,%
pravdivtsev_robust_2016,%
rodin_using_2018,%
rodin_constant-adiabaticity_2021,%
rodin_SOD_2019,%
sheberstov_excitation_2019,%
kharkov_singlet_2019,%
mamone_gcM2S_2020,%
bengs_generalised_2020,%
mamone_localized_2021,%
bengs_nuclear_2021,%
cavadini_slow_2005,%
cavadini_singlet_2008,%
sarkar_measurement_2008,%
ahuja_diffusion_2009,%
salvi_boosting_2012,%
buratto_exploring_2014,%
buratto_drug_2014,%
buratto_ligandprotein_2016,%
berner_sambadena_2019,%
mamone_localized_2021,
eills_singlet_2017,%
stevanato_pulse_2017,%
bengs_robust_2020,%
korenchan_31p_2021,%
kharkov_weak_2022,%
roy_initialization_2010,%
rodin_LLSAC_2020%
}%
\xspace}%
\newcommand{\citeSBR}{\cite{%
carravetta_symmetry_2000,%
levitt_symmetry-based_2007,%
levitt_symmetry_2008,%
brinkmann_symmetry_2001%
}\xspace}%
\newcommand{\citeAHT}{
\cite{haeberlen_coherent_1968,%
mansfield_symmetrized_1971,%
haeberlen_high_1976%
}%
\xspace}
\newcommand{\citeCompPulse}{
\cite{levitt_nmr_1979,%
levitt_compensation_1981,%
levitt_composite_1986,%
shaka_symmetric_1987,%
odedra_dual-compensated_2012%
}%
\xspace}
\newcommand{\citeFloquet}{%
\cite{%
leskes_floquet_2010,%
ivanov_floquet_2021%
}%
\xspace}

\makeatother
\begin{document}

\preprint{APS/123-QED}

\title{
Symmetry-Based Singlet-Triplet Excitation in Solution Nuclear Magnetic Resonance
}% Force line breaks with \\

\author{Mohamed Sabba}
\affiliation{School of Chemistry, University of Southampton, SO17 1BJ, UK}

\author{Nino Wili}
\affiliation{Interdisciplinary Nanoscience Center (iNANO) and Department of Chemistry, Aarhus University, Gustav Wieds Vej 14, DK-8000 Aarhus C, Denmark}

\author{Christian Bengs}
\affiliation{School of Chemistry, University of Southampton, SO17 1BJ, UK}

%\author{Laurynas Dagys}
%\affiliation{Department of Chemistry, University of Southampton, SO17 1BJ, UK

%\author{Maria Concistre}
%\affiliation{Department of Chemistry, University of Southampton, SO17 1BJ, UK}

\author{Lynda J. Brown}
\affiliation{School of Chemistry, University of Southampton, SO17 1BJ, UK}

\author{Malcolm H. Levitt}
 \email{mhl@soton.ac.uk}
 \affiliation{School of Chemistry, University of Southampton, SO17 1BJ, UK}

\date{\today}% It is always \today, today,
% comment out to disable highlight
%\renewcommand{\hl}[1]{#1}

\begin{abstract}
%\blue{
Coupled pairs of spin-1/2 nuclei support one singlet state and three triplet states. In many circumstances the nuclear singlet order, defined as the difference between the singlet population and the mean of the triplet populations, is a long-lived state which persists for a relatively long time in solution. Various methods have been proposed for generating singlet order, starting from nuclear magnetization. This requires the stimulation of singlet-to-triplet transitions by modulated radiofrequency fields. We show that a recently described pulse sequence, known as PulsePol (Schwartz \textit{et al.}, Science Advances, \textbf{4}, eaat8978 (2018)), is an efficient technique for converting magnetization into long-lived singlet order.  We show that the operation of this pulse sequence may be understood by adapting the theory of symmetry-based recoupling sequences in magic-angle-spinning solid-state NMR. The concept of riffling allows PulsePol to be interpreted using the theory of symmetry-based pulse sequences, and explains its robustness. This theory is used to derive a range of new pulse sequences for performing singlet-triplet excitation and conversion in solution NMR. Schemes for further enhancing the robustness of the transformations are demonstrated. 
%}
\end{abstract}

\maketitle

\section{\label{sec:intro}Introduction}
%----------------
\emph{Long-lived states} are configurations of nuclear spin state populations which, under suitable circumstances, are protected against important dissipation mechanisms and which therefore persist for unusually long times in solution~\citeLLS . The seminal example is the \emph{singlet order} of spin-1/2 pair systems, which is defined as the population imbalance between the spin $I=0$ nuclear singlet state of the spin pair, and the spin $I=1$ triplet manifold~\cite{MHL-LLS-ARPC_2012,Levitt-LLSS_2019}. Nuclear singlet order may be exceptionally long-lived, with decay time constants exceeding 1 hour in special cases~\cite{stevanato_nuclear_2015}. The phenomenon of long-lived nuclear spin order has been used for a variety of purposes in solution nuclear magnetic resonance (NMR), including the study of slow processes such as chemical exchange~\cite{sarkar_sequence_2007,bengs_nuclear_2021}, molecular transport~\cite{cavadini_slow_2005,cavadini_singlet_2008,sarkar_measurement_2008,ahuja_diffusion_2009}, and infrequent ligand binding to biomolecules~\cite{salvi_boosting_2012,buratto_exploring_2014,buratto_drug_2014,buratto_ligandprotein_2016}, as well as quantum information processing~\cite{roy_initialization_2010,rodin_LLSAC_2020}. The dynamics of nuclear singlet states is also central to the exploitation of parahydrogen spin order in hyperpolarized NMR experiments~\cite{bowers_parahydrogen_1987,pravica_net_1988,kadlecek_optimal_2010,eills_singlet_2017,stevanato_pulse_2017,bengs_robust_2020,dagys_low-frequency_2021,dagys_hyperpolarization_2022}. Singlet NMR has also been applied to imaging and \emph{in vivo} experiments~\cite{berner_sambadena_2019,mamone_localized_2021,devience_nuclear_2013,mamone_gcM2S_2020,huang_adaptable_2022,pileio_recycling_2013,eills_singlet-contrast_2021,pileio_real-space_2015,laustsen_hyperpolarized_2012,graafen_magnetic_2016,dumez_long-lived_2014,yang_multiple-targeting_2022}, and related techniques such as spectral editing \cite{pravdivtsev_vitro_2020,devience_nmr_2021} and low-field spectroscopy \cite{barskiy_nmr_2017,sjolander_13c-decoupled_2017,devience_homonuclear_2021,devience_homonuclear_2022}.

Several methods exist for converting nuclear magnetization into singlet order in the ``weak coupling" regime, meaning that the difference in the chemically shifted Larmor frequencies greatly exceeds the J-coupling between the members of the spin pair~\cite{carravetta_LFsinglet_2004,carravetta_HFsinglet_2004,sarkar_sequence_2007}. Methods for the ``near equivalent" and ``intermediate coupling" regimes (where the chemical shift frequency difference is weaker or comparable to the J-coupling), include the magnetization-to-singlet (M2S) pulse sequence~\cite{pileio_storage_2010,tayler_singlet_2011} and  variants such as gM2S~\cite{bengs_generalised_2020} and gc-M2S~\cite{mamone_gcM2S_2020}, the spin-lock-induced crossing (SLIC) method~\cite{devience_preparation_2013,devience_nuclear_2014,devience_probing_2016,devience_homonuclear_2021}, and slow passage through level anticrossings~\cite{pravdivtsev_robust_2016,rodin_using_2018}.

Recently, a new candidate sequence has emerged, namely the \emph{PulsePol} sequence, which was originally developed to implement electron-to-nuclear polarization transfer in the context of diamond nitrogen-vacancy magnetometry~\cite{schwartz_robust_2018,tratzmiller_pulsed_2021,tratzmiller_parallel_2021}. PulsePol is an attractively simple repeating sequence of six resonant pulses and four interpulse delays.  The PhD thesis of Tratzmiller~\cite{tratzmiller_pulsed_2021} reports numerical simulations in which PulsePol is used for magnetization-to-singlet conversion in the near-equivalent regime of high-field solution NMR. These simulations indicate that PulsePol could display significant advantages in robustness over some existing methods such as M2S and its variants. In this article we report the following: (i) the confirmation of Tratzmiller's proposal by experimental tests; (ii) the use of symmetry-based recoupling theory, as used in magic-angle-spinning solid-state NMR~\citeSBR, 
for elucidating the operation of this pulse sequence and predicting new ones; (iii) the PulsePol sequence and its variants may be used to excite singlet-triplet coherences;  (iv) the robustness of the singlet-triplet transformation may be enhanced further by using composite pulses. 

The PulsePol sequence was originally derived using average Hamiltonian theory with explicit solution of analytical equations~\cite{schwartz_robust_2018}. In this article we demonstrate an alternative theoretical treatment of PulsePol derived from the principles of symmetry-based recoupling in magic-angle-spinning solid-state NMR~\citeSBR.
This theoretical relationship is surprising since singlet-to-triplet conversion in solution NMR appears to be remote from recoupling in rotating solids. Nevertheless, as shown below, the problem of singlet-triplet conversion may be analysed in a time-dependent interaction frame in which the nuclear spin operators acquire a periodic time-dependence through the action of the scalar spin-spin coupling. The time-dependent spin operators in the interaction frame may be treated in similar fashion to the anisotropic spin interactions in rotating solids, in which case the periodic time-dependence is induced by the mechanical rotation of the sample. In both contexts, selection rules for the average Hamiltonian terms may be engineered by imposing symmetry constraints on the applied pulse sequences.

One common implementation of PulsePol corresponds to the pulse sequence symmetry designated \Rsymm431, using the notation developed for symmetry-based recoupling~\citeSBR. 
As shown below, the spin dynamical selection rules associated with \Rsymm431 symmetry explain the main properties of the PulsePol sequence. Furthermore this description immediately predicts the existence of many other sequences with similar properties. Some of these novel sequences are demonstrated below. 

PulsePol deviates from the standard construction procedure for symmetry-based recoupling sequences in solids. The deviation is subtle but invests PulsePol with improved robustness. Incorporating composite pulses can increase the robustness further. 

%---------------------
\section{Theory}
%-----
\subsection{Spin Hamiltonian}
%------
The rotating-frame spin Hamiltonian for a homonuclear 2-spin-1/2 system in high-field solution NMR may be written as
\begin{equation}\label{eq:H}
    H(t) = H_\mathrm{CS} +H_{J} + \Hrf(t),
\end{equation}
where the chemical shift Hamiltonian is given by
\begin{equation}\label{eq:HCS}
    H_\mathrm{CS} = H_{\Sigma}+ H_{\Delta}
\end{equation}
and the individual Hamiltonian terms are:
\begin{equation}\label{eq:Hterms}
\begin{aligned}
H_{\Sigma}&=\hf\wS (I_{1z}+I_{2z}),
\\
H_{\Delta}&=\hf\wD (I_{1z}-I_{2z}), 
\\
H_{J}&= \wJ \boldI_1\cdot\boldI_2.
%+ \Hrf(t),
\end{aligned}
\end{equation}
Here, $\wS$ is the sum of the chemically shifted resonance offsets for the two spins, $\wD$ is their difference, and $\wJ=2\pi J$ is the scalar spin-spin coupling ($J$-coupling). 

The interaction of the spin pair with resonant radiofrequency fields is represented by the Hamiltonian term $\Hrf(t)$. The rotating-frame Hamiltonian for the interaction of the nuclei with a resonant time-dependent field is given by
\begin{equation}
\Hrf(t)=\wnut(t) \left\{
\cos\phi(t)(I_{1x}+I_{2x})
+\sin\phi(t)(I_{1y}+I_{2y})
\right\},
\end{equation}
where the nutation frequency $\wnut$ is proportional to the radiofrequency field amplitude.

The terms $H_{\Sigma}$, $H_J$ and $\Hrf$ all mutually commute. The term $H_\Delta$, on other hand, commutes in general with neither $H_J$ nor $\Hrf$. We consider here the case of ``near-equivalent" spin pairs~\cite{tayler_singlet_2011,pileio_storage_2010,devience_preparation_2013}, for which $\abs{\wD}\ll\abs{\wJ}$. In this case, the term $H_\Delta$ may be treated as a perturbation of the dominant terms $H_J$ and $H_\mathrm{rf}$. 

%-----
\subsection{Propagators}
%------
The propagator $U_\Lambda(t)$ generated by a Hamiltonian term $H_\Lambda$ is a unitary time-dependent operator solving the differential equation
\begin{align}\label{eq:U}
    \frac{\mathrm{d}}{\mathrm{d}t}
    U_\Lambda(t)
    &=
    - i H_\Lambda(t) U_\Lambda(t)
\end{align}
with the boundary condition $U_\Lambda(0)=1$. Since $\Hrf$ and $H_J$ commute, the propagator $U(t)$ under the total Hamiltonian of equation~\ref{eq:H} may be written as follows:
\begin{equation}\label{eq:U}
U(t)=
    U_J(t)
    U_\mathrm{rf}(t)
\widetilde{U}_\mathrm{CS}(t),
\end{equation}
where the propagator $\widetilde{U}_\mathrm{CS}(t)$ solves the differential equation
\begin{align}\label{eq:UtildeCS}
    \frac{\mathrm{d}}{\mathrm{d}t}
   \widetilde{U}_\mathrm{CS}(t)
    &=
    - i 
    \widetilde{H}_\mathrm{CS}(t)
    \widetilde{U}_\mathrm{CS}(t)
\end{align}
with the boundary condition $\widetilde{U}_\mathrm{CS}(0)=1$. 
The interaction-frame chemical shift Hamiltonian $\widetilde{H}_\mathrm{CS}(t)$ is defined as follows:
\begin{equation}\label{eq:HtildeCS}
    \widetilde{H}_\mathrm{CS}(t)
    =
    U_\mathrm{rf}(t)^\dagger
        U_J(t)^\dagger
        H_\mathrm{CS} 
        U_J(t)
        U_\mathrm{rf}(t).
\end{equation}
Equation~\ref{eq:HtildeCS} shows that the chemical shift terms acquire a double modulation in the interaction frame: first from the action of the J-coupling, and secondly from the action of the applied rf field. 

Since the J-coupling is time-independent, the propagator $U_J$ has the following form:
\begin{align}
    U_J(t) = \exp\{-i H_J t\}
    = \exp\{ -i \wJ t \boldI_1\cdot\boldI_2 \}.
\end{align}
The singlet and triplet states of the spin-1/2 pair are defined as follows:
\begin{align}
    \ket{S_0} &= 
        2^{-1/2}(\ket{\alpha\beta}-\ket{\beta\alpha}),
\nonumber\\
    \ket{T_{+1}} &= 
        \ket{\alpha\alpha},
\nonumber\\
    \ket{T_0} &= 
        2^{-1/2}(\ket{\alpha\beta}+\ket{\beta\alpha}),
\nonumber\\
    \ket{T_{-1}} &= 
        \ket{\beta\beta}.
\end{align}
Since the singlet and triplet states are eigenstates of $H_J$, with eigenvalues $-3\omega_J/4$ and $+\omega_J/4$ respectively, the propagator $U_J$ may be written as follows:
\begin{align}\label{eq:UJinST}
     U_J(t) = 
        & \exp\{+i\tfrac{3}{4}\omega_J t\}
            \ket{S_0}\bra{S_0}
\nonumber\\
    & + \exp\{-i\tfrac{1}{4}\omega_J t\}
        \sum_M \ket{T_M}\bra{T_M}.
\end{align}

The rf propagator $U_{\rm rf}(t)$ corresponds to a time-dependent rotation in three-dimensional space, described by three Euler angles:
\begin{equation}\label{eq:Urf}
\begin{aligned}
U_{\rm rf}(t)=&R(\Omega_{\rm rf}(t))
\\
=&R_{z}(\alpha_{\rm rf}(t))R_{y}(\beta_{\rm rf}(t))R_{z}(\gamma_{\rm rf}(t)),
\end{aligned}
\end{equation}
with
\begin{equation}
\begin{aligned}
R_{\chi}(\theta)=\exp\{-\iu \theta I_{\chi}\}.
\end{aligned}
\end{equation}
The action of the modulated radiofrequency field on the spin system may therefore be described in terms of a time-dependent set of three Euler angles $\Omega_{\rm rf}(t)=\{\alpha_{\rm rf}(t),\beta_{\rm rf}(t),\gamma_{\rm rf}(t)\}$.

In general, it is possible to modulate the amplitude $\wnut(t)$ and phase $\phi(t)$ of the rf field in time, in order to generate any desired trajectory of Euler angles $\Omega_{\rm rf}(t)$.

%-----------
\subsection{Spherical Tensor Operators}
%-----------
It is convenient to define two spherical tensor spin operators of rank-1, denoted $\mathds{T}_1^g$ and  $\mathds{T}_1^u$, where the superscripts denote their parity under exchange of the two spin-1/2 particles:
\begin{align}
    (12)\mathds{T}_{1m}^g     (12)^\dagger 
    &=\mathds{T}_{1m}^g,
\nonumber\\
    (12)\mathds{T}_{1m}^u     (12)^\dagger 
    &=-\mathds{T}_{1m}^u,
\end{align}
where $m\in\{+1,0,-1\}$ and $(12)$ denotes the particle exchange operator. The \emph{gerade} spherical tensor operator is constructed from the total angular momentum and shift operators for the spin system:
\begin{align}
  \mathds{T}_{1\,+1}^g
    &= -2^{-1/2}(I_1^+ + I_2^+ ),
\nonumber\\
  \mathds{T}_{1\,0}^g
    &= I_{1z} + I_{2z},
\nonumber\\  \mathds{T}_{1\,-1}^g
    &= 2^{-1/2}(I_1^- + I_2^- ).
\end{align}
The \emph{ungerade} spherical tensor operator of rank-1 plays a prominent role in the current theory. It has the following components:
\begin{align}
  \mathds{T}_{1\,+1}^u
    &= \ket{T_{+1}}\bra{S_0},
\nonumber\\
  \mathds{T}_{1\,0}^u
    &= \ket{T_{0}}\bra{S_0},
\nonumber\\  \mathds{T}_{1\,-1}^u
    &= \ket{T_{-1}}\bra{S_0}.
\end{align}
Each component is given by a shift operator between the singlet state and one of the three triplet states. The adjoint operators are given by
\begin{align}
  \mathds{T}_{1\,+1}^{u\dagger}
    &= \ket{S_0}\bra{T_{+1}},
\nonumber\\
  \mathds{T}_{1\,0}^{u\dagger}
    &= \ket{S_0}\bra{T_{0}},
\nonumber\\  \mathds{T}_{1\,-1}^{u\dagger}
    &= \ket{S_0}\bra{T_{-1}}.
\end{align}

Both sets of operators $\mathds{T}_{1}^{g}$ and $\mathds{T}_{1}^{u}$
transform irreducibly under the three-dimensional rotation group:
\begin{equation}\label{eq:ISTOrots}
\begin{aligned}
R(\Omega)\mathds{T}^{g}_{1\mu}R^{\dagger}(\Omega)
&=
\sum_{\mu'=-1}^{+1}\mathds{T}^{g}_{1\mu'}\mathcal{D}^{1}_{\mu'\mu}(\Omega),
\\
R(\Omega)\mathds{T}^{u}_{1\mu}R^{\dagger}(\Omega)
&=
\sum_{\mu'=-1}^{+1}\mathds{T}^{u}_{1\mu'}\mathcal{D}^{1}_{\mu'\mu}(\Omega).
\\
\end{aligned}
\end{equation}
Here, $\mathcal{D}^{\lambda}_{\mu'\mu}(\Omega)$ represents an element of the rank-$\lambda$ Wigner rotation matrix~\cite{VMK_1988}.

The \emph{gerade} spherical tensor operator $\mathds{T}_1^g$ obeys the standard relationship between its components under the adjoint transformation~\cite{VMK_1988}:
\begin{equation}
   \mathds{T}_{1\mu}^{g\dagger}=
   (-1)^\mu 
   \mathds{T}_{1\,-\mu}^{g}.
\end{equation}
However, the analogous relationship does \emph{not} apply to the components of the \emph{ungerade} spherical tensor operator $\mathds{T}_1^u$.

%-----------
\subsection{Interaction frame Hamiltonian}
%-----------
The chemical shift Hamiltonian terms, given in equation~\ref{eq:Hterms}, may be written in terms of the $m=0$ spherical tensor operator components as follows:
\begin{align}
H_\Sigma &=
    \hf\wS \mathds{T}_{1\,0}^g,
\nonumber\\
H_\Delta &=
    \hf\wD \big(
    \mathds{T}_{1\,0}^u
    +\mathds{T}_{1\,0}^{u\dagger}
    \big).
\end{align}
From equation~\ref{eq:UJinST}, these operators transform as follows under the propagator $U_J$:
\begin{align}
U_J^\dagger(t)
H_\Sigma U_J(t)
&=\hf\wS \mathds{T}_{1\,0}^g,
\nonumber\\
U_J(t)^\dagger
H_\Delta U_J(t)
&=\hf\wD \big(
    \mathds{T}_{1\,0}^u \exp\{-i\omega_J t\}
    +\mathds{T}_{1\,0}^{u\dagger}
    \exp\{+i\omega_J t\}
    \big).
\end{align}
This may be combined with equations~\ref{eq:HtildeCS}, \ref{eq:Urf} and \ref{eq:ISTOrots} to obtain the following expression for the interaction-frame chemical shift Hamiltonian: 
\begin{equation}
\widetilde{H}_\mathrm{CS}(t)=
    \sum_{m=-1}^{+1} {\ }
    \sum_{\mu=-1}^{+1}
    \widetilde{H}_{1m1\mu}(t),
\end{equation}
where each term has the form
\begin{equation}\label{eq:Htildemmu}
\widetilde{H}_{1m1\mu}(t)
=
\omega_{1m1\mu}
d^1_{\mu 0}\big(-\beta_\mathrm{rf}(t)\big)
\exp\{
i \big(
 m \omega_J + \mu \gamma_\mathrm{rf}(t)
\big)
\}
Q_{1m1\mu}
\end{equation}
and $d^1_{\mu 0}(\beta)$ is an element of the rank-1 reduced Wigner matrix. The amplitudes $\omega_{1m1\mu}$ and spin operators $Q_{1m1\mu}$ take the following values:
\begin{align}\label{eq:wmmuandQmmu}
    \omega_{1\,+1\,1\mu}   &= 
        \hf\omega_\Delta
    {\ },\quad&
    Q_{1\,+1\,1\mu} &
        =\mathds{T}_{1\,\mu}^u,
\nonumber\\
    \omega_{1\,0\,1\mu}   &= \hf\omega_\Sigma
    {\ },\quad&
    Q_{1\,0\,1\mu} &=\mathds{T}_{1\,\mu}^g,
\nonumber\\
    \omega_{1\,-1\,1\mu}   &= \hf\omega_\Delta
    {\ },\quad&
    Q_{1\,-1\,1\mu} &=(-1)^\mu\mathds{T}_{1\,-\mu}^{u\dagger},
\end{align}
where $\mu\in\{+1,0,-1\}$. Note that the singlet-triplet excitation terms have quantum number $m=\pm1$, while the resonance offset term has $m=0$. 

For the terms $\omega_{\ell m \lambda \mu}$ and $Q_{\ell m \lambda \mu}$ above, the rank of the interaction under rotations of the spins is specified as $\lambda=1$. The 
``pseudo-space-rank" $\ell=1$, on the other hand, has no physical meaning and is introduced to establish a correspondence with the notation used in magic-angle-spinning solid-state NMR~\citeSBR.

%-----------
\begin{figure}[tbh]
\includegraphics[width=0.5\textwidth]{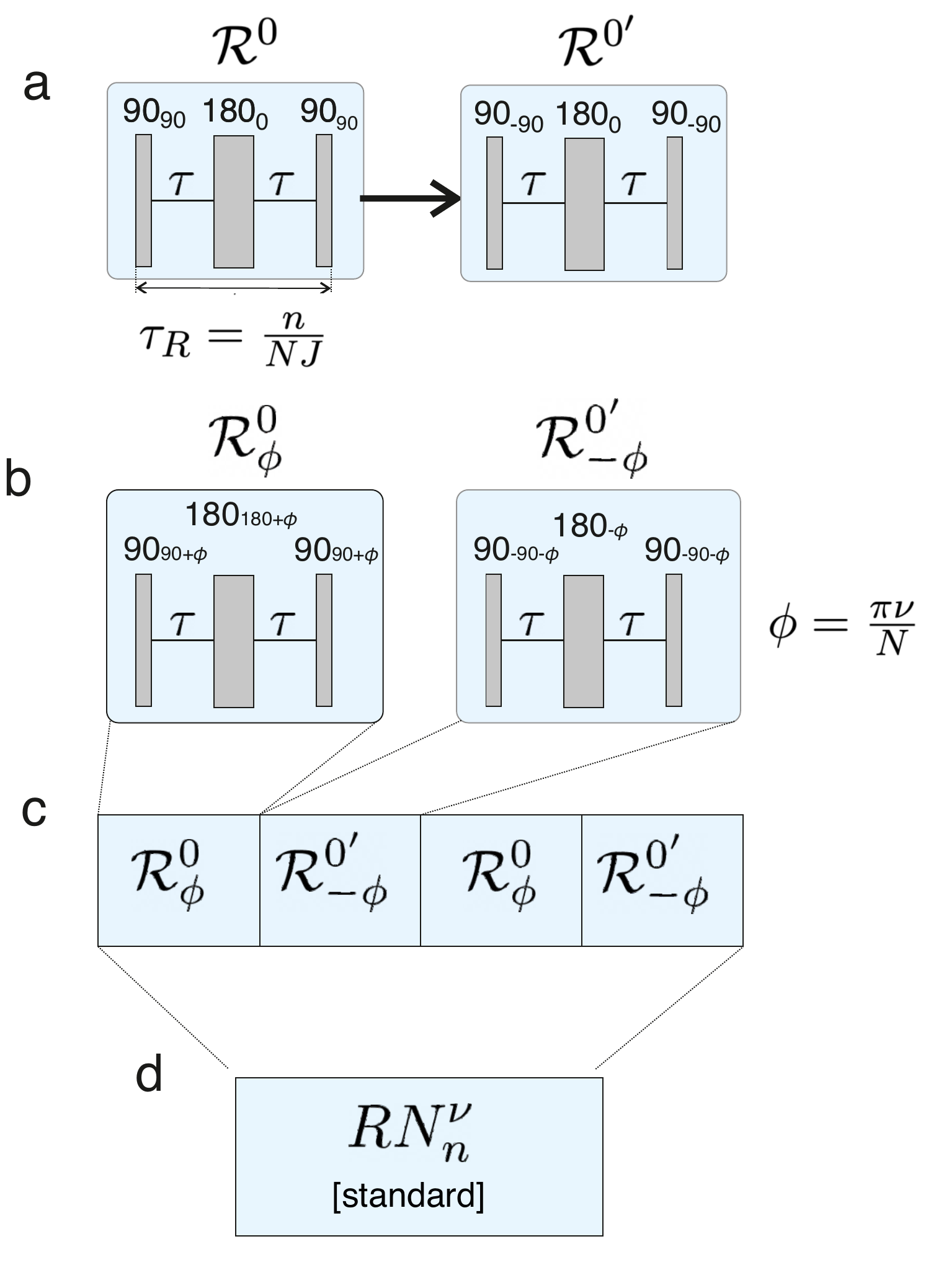}
\caption{
\label{fig:Rseqstandard}
Standard implementation of a \RNnnu sequence for singlet-triplet conversion. (a) A basic $R$-element denoted $\mathcal{R}^0$ is selected. This element induces a rotation about the rotating-frame x-axis through an odd multiple of $\pi$. In the current case, the element $\mathcal{R}^0$ is given by the composite pulse $\mathrm{90_{90} 180_0 90_{90}}$ with delays $\tau$ between the pulses, such that its overall duration is $\tau_R=n/(NJ)$. The conjugate sequence $\mathcal{R}^{0'}$ is generated from $\mathcal{R}^{0}$ by a change in sign of all phases. (b) The sequence $\mathcal{R}^0$ is given a phase shift of $+\phi$, while the sequence $\mathcal{R}^{0'}$ is given a phase shift of $-\phi$, where $\phi=\pi\nu/N$. (c) The pair of sequences $(\mathcal{R}^0)_\phi$ and $(\mathcal{R}^{0'})_{-\phi}$ is repeated $N/2$ times, to give the standard implementation of a \RNnnu sequence (d). 
}
\end{figure}
\subsection{Symmetry-Based Sequences}
%-----------
Symmetry-based pulse sequences~\citeSBR 
were originally developed for magic-angle-spinning solid-state NMR, where the sample is rotated mechanically with the angular frequency $\omega_r$, such that its rotational period is given by $\tau_r=\abs{2\pi/\omega_r}$. In the current case of singlet-triplet excitation in solution NMR, the J-coupling plays the role of the mechanical rotation. The relevant period is therefore given by $\tau_J = \abs{2\pi/\omega_J} = \abs{J^{-1}}$. 

In the current context, a sequence with \RNnnu symmetry is defined by the following time-symmetry relationship of the rf Euler angles $\beta_\mathrm{rf}(t)$ and $\gamma_\mathrm{rf}(t)$, which applies for arbitrary time points $t$~\citeSBR:
\begin{align}\label{eq:EulerRsymm}
\beta_\mathrm{rf}(t+\frac{n\tau_J}{N}) &= \beta_\mathrm{rf}(t) \pm \pi,
\nonumber\\
\gamma_\mathrm{rf}(t+\frac{n\tau_J}{N}) &= \gamma_\mathrm{rf}(t) 
-\frac{2\pi\nu}{N}.
\end{align}
A complete \RNnnu sequence has duration $T=n\tau_J$, and is cyclic, in the sense that the net rotation induced by the rf field over the complete sequence is through an even multiple of $\pi$. 

The \emph{symmetry numbers} $N$, $n$ and $\nu$ take integer values. In the case of \RNnnu sequences, $N$ must be even, while $n$ and $\nu$ are unconstrained. As discussed below, the symmetry numbers define the selection rules for the spin dynamics under the pulse sequence. 

The \RNnnu Euler angle symmetries in equation~\ref{eq:EulerRsymm} do not define the pulse sequence uniquely. Nevertheless, there is a standard procedure~\citeSBR
for generating these Euler angle symmetries, which is sketched in figure~\ref{fig:Rseqstandard}. The procedure is as follows: 
\begin{itemize}
\item Select a rf pulse sequence known as a \emph{basic R-element}, designated $\mathcal{R}^0$. This sequence may be arbitrarily complex, but must induce a net rotation of the resonant spins by an odd multiple of $\pi$ about the rotating-frame x-axis. If the duration of the basic element $\mathcal{R}^0$ is denoted $\tau_R$, this implies the condition
\begin{equation}
    U_\mathrm{rf}(\tau_R) =
     R_x(p\pi),
\end{equation}
where $p$ is an odd integer.
\item The duration of the basic element $\tau_R$ is given by $\tau_R=(n/N) J^{-1}$, where $n$ and $N$ are the symmetry numbers of the \RNnnu sequence. 
\item Reverse the sign of all phases in $\mathcal{R}^0$. This leads to the \emph{conjugate element} designated $\mathcal{R}^{0'}$.
\item Give all components of the basic element $\mathcal{R}^0$ a phase shift of $+\pi\nu/N$. This gives the phase-shifted basic element, denoted $\mathcal{R}^{0}_{+\pi\nu/N}$.
\item Give all components of the conjugate element $\mathcal{R}^{0'}$ a phase shift of $-\pi\nu/N$. This gives the element  $\mathcal{R}^{0'}_{-\pi\nu/N}$.
\item The complete \RNnnu sequence is composed of $N/2$ repeats of the element pair, as follows:
\begin{equation}\label{eq:RNnnustand}
\RNnnu =
    \{\mathcal{R}^{0}_{+\pi\nu/N}\mathcal{R}^{0'}_{-\pi\nu/N}\}^{N/2}.
\end{equation}
\end{itemize}
The complete \RNnnu sequence has an overall duration of 
\begin{equation}
\begin{aligned}
T=N\tau_R = n J^{-1}.
\end{aligned}
\end{equation}

%-----------
\subsection{Selection Rules}
%-----------
The propagator for a complete \RNnnu sequence is given from equation~\ref{eq:U} by
\begin{equation}
U(T)=
    U_J(T)
    U_\mathrm{rf}(T)
\widetilde{U}_\mathrm{CS}(T).
\end{equation}
From the definition of a \RNnnu sequence, the complete sequence propagators $U_J(T)$ and $U_\mathrm{rf}(T)$ are both proportional to the unity operator and may be ignored. The operator $\widetilde{U}_\mathrm{CS}(T)$ corresponds to propagation under a time-independent effective Hamiltonian:
\begin{equation}
 \widetilde{U}_\mathrm{CS}(T) =
 \exp\{
 -i
 \overline{H}_\mathrm{CS} T
 \}.
\end{equation}
In the near-equivalence limit ($\abs{\omega_J}\gg\abs{\omega_\Delta}, \abs{\omega_\Sigma}$), the effective Hamiltonian  $\overline{H}_\mathrm{CS}$ may be approximated by the first term in a Magnus expansion~\citeAHT:
\begin{align}
\overline{H}_\mathrm{CS}
    &\simeq
\overline{H}_\mathrm{CS}^{(1)},
\end{align}
where
\begin{equation}
   \overline{H}_\mathrm{CS}^{(1)} 
   =
   \sum_{m=-1}^{+1}
   \sum_{\mu=-1}^{+1}
   \overline{H}_{1m1\mu}^{(1)}. 
\end{equation}
In common with many recent papers~\citeSBR, 
this article uses a numbering of the Magnus expansion terms which differs from the older literature\citeAHT
by one.

The individual average Hamiltonian terms are given by
\begin{equation}
    \overline{H}_{1m1\mu}^{(1)}
    =
    T^{-1}\int_0^T
        \widetilde{H}_{1m1\mu}(t)
        {\ }dt,
\end{equation}
where the interaction frame terms $\widetilde{H}_{1m1\mu}(t)$ are given in equation~\ref{eq:Htildemmu}.

The Euler angle symmetries in equation~\ref{eq:EulerRsymm} lead to the following selection rules for the first-order average Hamiltonian terms of \RNnnu sequences~\citeSBR:
\begin{equation}
\label{eq:RSrule}
\begin{aligned}
\overline{H}^{(1)}_{\ell m\lambda\mu}(t_{0})=0 \quad {\rm if}\;mn-\mu\nu\neq\frac{N}{2}Z_{\lambda},
\end{aligned}
\end{equation}
where $Z_{\lambda}$ is any integer with the same parity as $\lambda$. This selection rule may be visualised by a diagrammatic procedure~\cite{levitt_symmetry-based_2007,levitt_symmetry_2008}.  

In the current case, $\lambda=1$ for all relevant interactions, so that $Z_{\lambda}$ is any odd integer. Hamiltonian components for which $mn-\mu\nu$ is an odd multiple of $N/2$ are symmetry-allowed and may contribute to the effective Hamiltonian. A symmetry-allowed term with quantum numbers $\{m,\mu\}$ and ranks $\ell=\lambda=1$ is given in general by
\begin{equation}
  \overline{H}^{(1)}_{1 m 1\mu}  
  =
  \kappa_{1 m 1\mu} 
  \omega_{1 m 1\mu} 
  Q_{1 m 1\mu},
\end{equation}
where the amplitudes $\omega_{1 m 1\mu}$ and spin operators $Q_{1 m 1\mu}$ are given in equation~\ref{eq:wmmuandQmmu}. 

The scaling factor $\kappa_{\ell m\lambda\mu}$ of a symmetry-allowed term is given by
\begin{equation}
\begin{aligned}
&\kappa_{\ell m\lambda\mu}=
\exp(-\iu\mu\frac{\pi\nu}{N})K_{m\lambda\mu},
\end{aligned}
\end{equation}
where $K_{m\lambda\mu}$ is defined with respect to the basic element $\mathcal{R}^0$:
\begin{equation}
\begin{aligned}
K_{m\lambda\mu}=
\tau^{-1}_R\int_{0}^{\tau_R} d^{\lambda}_{\mu0}(-\beta^{0}_{\rm rf}(t))
\exp\{i(\mu \gamma^{0}_{\rm rf}(t)+m \wJ t)\}dt.
\end{aligned}
\end{equation}
Here $\beta^{0}_{\rm rf}$ and $\gamma^{0}_{\rm rf}$ represent the Euler angles describing the rotation induced by the rf field under the basic element~\citeSBR.

Symmetry-based pulse sequences are designed by selecting combinations of symmetry numbers $N$, $n$ and $\nu$ such that all desirable average Hamiltonian terms $\overline{H}^{(1)}_{\ell m \lambda\mu}$ are symmetry-allowed while all undesirable terms are symmetry-forbidden. In most cases, the basic element $\mathcal{R}^0$ is selected such that the scaling factors $\kappa_{\ell m\lambda\mu}$ are maximised for the desirable symmetry-allowed terms.

\begin{figure}[tb]
\centering
\includegraphics[trim={0 0 20cm 0},clip,width=0.8\columnwidth]{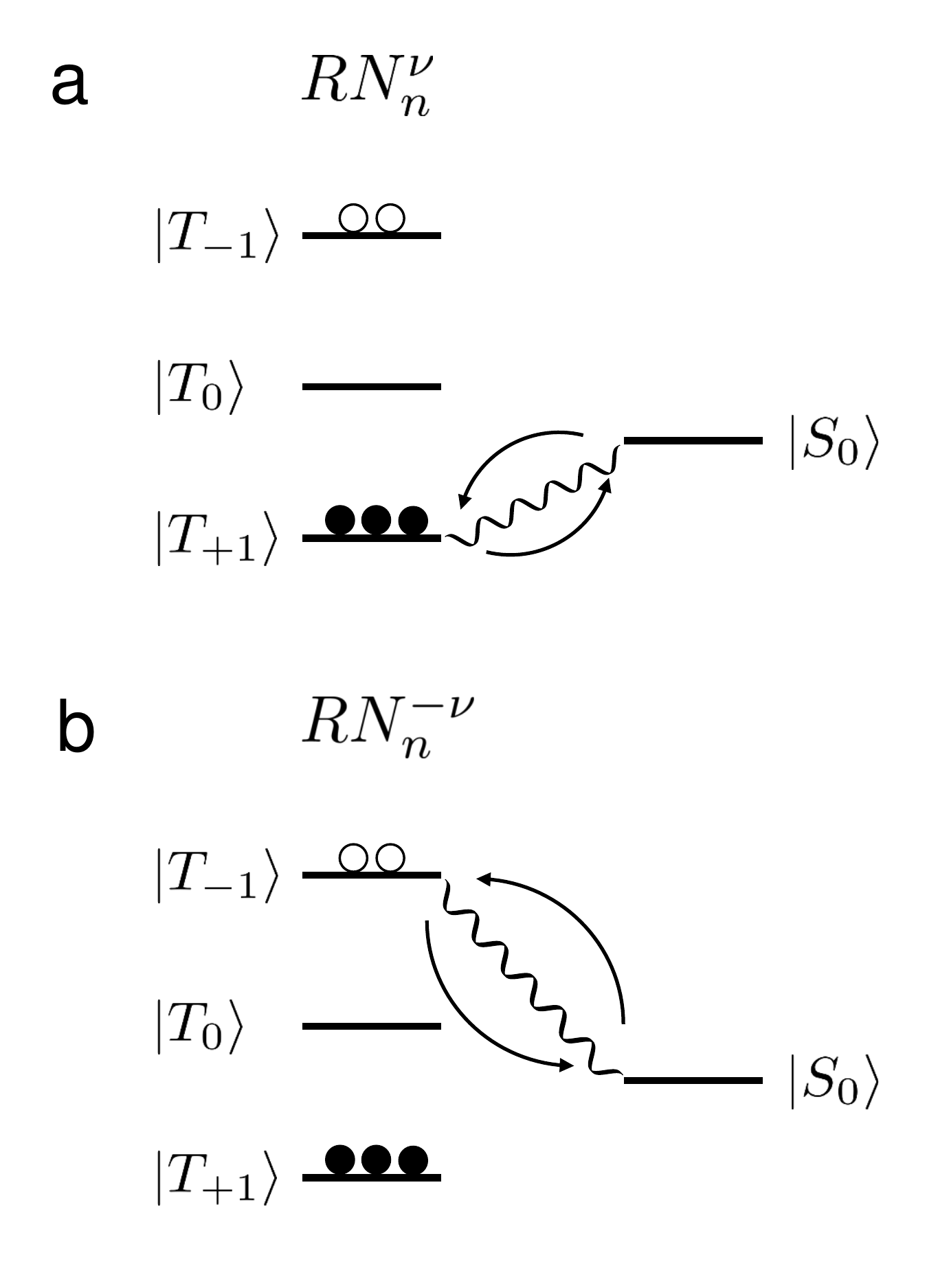}
\caption{\label{fig:transitions} Energy levels and approximate eigenstates of a J-coupled two-spin-1/2 system in the near-equivalence limit. (a) A \RNnnu sequence, with symmetry numbers chosen to select terms $\{m,\mu\}=\{\pm1,\pm1\}$ and suppress all others, induces a transition between the $\ket{S_0}$ and $\ket{T_{+1}}$ states. Suitable symmetries are given in table~\ref{tab:Symmetries}. One example is \Rsymm43{+1}. (b) If the symmetry number $\nu$ is changed in sign, average Hamiltonian terms with quantum numbers $\{m,\mu\}=\{\pm1,\mp1\}$ are selected. In this case there is selective excitation of the transition between the $\ket{S_0}$ and $\ket{T_{-1}}$ states. One example is \Rsymm43{-1}.
%\MHLnote{We have to be careful not to give the impression that a positive $\nu$ excites the lower transition, while a negative $\nu$ excites the upper transition. This is not always true (see table~\ref{tab:Symmetries}).
%\MHLnote{A small point, but make the (a), (b) labelling sans-serif, with the same font as the other figures.}
}
\end{figure}

 \begin{table}[tb]
\renewcommand{\arraystretch}{1.5}
\vspace{5mm}
\begin{tabular}{|l|r| } 
 \hline
 \RNnnu & $\kappa_{1111}$
  \\[3pt]\hline\hline
\Rsymm41{-1} & $-0.264$
 \\[3pt]\hline
\Rsymm431 & $-0.512$
 \\[3pt]\hline
\Rsymm45{-1} & $0.307$
 \\[3pt]\hline
\Rsymm47{1} & $0.038$
 \\[3pt]\hline
\Rsymm49{-1} & $-0.029$
 \\[3pt]\hline
\Rsymm61{-2} & $-0.104$
 \\[3pt]\hline
\Rsymm65{2} & $-0.291$
 \\[3pt]\hline
\Rsymm67{-2} & $0.360$
 \\[3pt]\hline
\Rsymm68{-1} & $0.253$
 \\[3pt]\hline
\Rsymm6{10}{1} & $0.068$
 \\[3pt]\hline
\end{tabular}
\qquad
\begin{tabular}{|l|r| } 
 \hline
 \RNnnu & $\kappa_{1111}$
  \\[3pt]\hline\hline
\Rsymm81{-3} & $-0.137$
 \\[3pt]\hline
\Rsymm83{-1} & $-0.371$
 \\[3pt]\hline
\Rsymm85{1} & $-0.498$
 \\[3pt]\hline
\Rsymm87{3} & $-0.495$
 \\[3pt]\hline
\Rsymm89{-3} & $0.385$
 \\[3pt]\hline
\Rsymm{10}1{-4} & $-0.110$
 \\[3pt]\hline
\Rsymm{10}2{-3} & $-0.215$
 \\[3pt]\hline
\Rsymm{10}3{-2} & $-0.309$
 \\[3pt]\hline
\Rsymm{10}4{-1} & $-0.389$
 \\[3pt]\hline
\Rsymm{10}6{1} & $-0.491$
 \\[3pt]\hline
\Rsymm{10}7{2} & $-0.511$
 \\[3pt]\hline
\end{tabular}
\caption{A selection of \RNnnu symmetries that are appropriate for symmetry-based singlet-triplet conversion in solution NMR. These symmetries select $\overline{H}^{(1)}_{\ell m \lambda\mu}$ terms with quantum numbers $\{\ell,m,\lambda,\mu\}$ given by $\{1,\pm1,1,\pm1\}$. Changing the sign of $\nu$ selects the terms $\{1,\pm1,1,\mp1\}$ instead. Scaling factors $\kappa_{1111}$ are given for the basic R-element in equation~\ref{eq:R0}, in the limit of radiofrequency pulses with negligible duration. 
\label{tab:Symmetries}
%\CBnote{these scaling factors assume magic angle spinning.}
%\MHLnote{Well spotted! I think I have corrected them.}
}
\end{table}
\subsection{Transition-selective singlet-triplet excitation}

Table~\ref{tab:Symmetries} shows some sets of symmetry numbers $\{N,n,\nu\}$ under which the average Hamiltonian terms with quantum numbers $\{\ell,m,\lambda,\mu\}$ $=$ $\{1,\pm1, 1,\pm1\}$ are symmetry-allowed, while all other terms are symmetry-forbidden and are suppressed in the average Hamiltonian. In particular, all resonance-offset terms, which have $m=0$, are symmetry-forbidden in the first-order average Hamiltonian, for the symmetries in table~\ref{tab:Symmetries}.

For example, consider the symmetry \Rsymm431. The term $\{\ell,m,\lambda,\mu\}=\{1,1,1,1\}$ is symmetry-allowed since the expression $nm-\nu\mu$ evaluates to $3\times1 - 1\times 1 = 2$, which is an odd multiple of $N/2=2$. The term $\{\ell,m,\lambda,\mu\}=\{1,1,1,-1\}$, on the other hand, is symmetry-forbidden, since $nm-\nu\mu$ evaluates to $3\times1 - 1\times (-1) = 4$, which is an \emph{even} multiple of $2$. Similarly, the resonance-offset term  $\{\ell,m,\lambda,\mu\}=\{1,0,1,-1\}$ is symmetry-forbidden, since $nm-\nu\mu$ evaluates to $3\times0 - 1\times (-1) = 1$, which is not an integer multiple of $2$.

All symmetries in table~\ref{tab:Symmetries} select Hamiltonian components with quantum numbers $\{\ell,m,\lambda,\mu\}=\{1,\pm1, 1,\pm1\}$, while suppressing all other terms. In this case the first-order average Hamiltonian is given through equations~\ref{eq:wmmuandQmmu} by 
\begin{align}
  \overline{H}_\mathrm{CS}^{(1)}
  &=
 \kappa_{1\,+1 1\,+1} 
  \omega_{1\,+1 1\,+1}
  Q_{1\,+1 1\,+1}
 \nonumber\\
  & +
\kappa_{1\,-1 1\,-1} 
  \omega_{1\,-1 1\,-1}
  Q_{1\,-1 1\,-1}
\nonumber\\
  &=
 \frac{1}{2}\wD\big\{
\kappa_{1+11+1}\mathds{T}^{u}_{1\,+1}
+(
\kappa_{1+11+1}\mathds{T}^{u}_{1\,+1}
)^\dagger
\big\}.
\end{align}
The first-order average Hamiltonian therefore generates a selective rotation of the transition between the singlet state $\ket{S_0}$ and the lower triplet state $\ket{T_{+1}}$, as shown in figure~\ref{fig:transitions}(a):
\begin{equation}
\label{eq:H1CS}
\begin{aligned}
\overline{H}^{(1)}_{\rm CS}
=\hf\wSTnut
\big(
e^{-i{\phi_\mathrm{ST}}}
\ket{S_0}\bra{T_{+1}}
+e^{+i{\phi_\mathrm{ST}}}
\ket{T_{+1}}\bra{S_0}
\big)
\end{aligned}
\end{equation}
The singlet-triplet nutation frequency and phase depend upon the scaling factors as follows
\begin{equation}
\begin{aligned}
\label{eq:wSTnut}
\wSTnut
=\wD\vert\kappa_{1+11+1}\vert
=\wD\vert\kappa_{1-11-1}\vert,
\end{aligned}
\end{equation}
\begin{equation}
\label{eq:wSTphase}
\begin{aligned}
\phi_\mathrm{ST}=
{\rm arg}(\kappa_{1-11-1})
={\rm arg}(-\kappa^{*}_{1111}).
\end{aligned}
\end{equation}

If a set of symmetry numbers $\{N,n,\nu\}$ selects the terms $\{\ell,m,\lambda,\mu\}=\{1,\pm1, 1,\pm1\}$, then the set of symmetry numbers $\{N,n,-\nu\}$ selects the terms $\{\ell,m,\lambda,\mu\}=\{1,\pm1, 1,\mp1\}$. As indicated in figure~\ref{fig:transitions}b, the change in sign of $\nu$ leads to a selective rotation of the singlet state and the \emph{upper} triplet state.

In either case the dynamics of the system may be described by a two-level treatment. Define the single-transition operators~\cite{wokaun_selective_1977,vega_fictitious_1978} for the transitions between the singlet state and the outer triplet states:
\begin{align}
I_x^{\mathrm{ST}(\pm)} &= 
   \frac{1}{2}\big(
   \ket{T_{\pm1}}\bra{S_0}
   + \ket{S_0}\bra{T_{\pm1}}
   \big),
\nonumber\\
I_y^{\mathrm{ST}(\pm)} &= 
   \frac{1}{2i}\big(
   \ket{T_{\pm1}}\bra{S_0}
   - \ket{S_0}\bra{T_{\pm1}}
   \big),
\nonumber\\
I_z^{\mathrm{ST}(\pm)} &= 
   \frac{1}{2}\big(
   \ket{T_{\pm1}}\bra{T_{\pm1}}
   - \ket{S_0}\bra{S_0}
   \big).
\end{align}
These operators have the cyclic commutation relationships~\cite{wokaun_selective_1977,vega_fictitious_1978}:
\begin{equation}
\label{eq:STCyComm}
    \big[
    I_x^{\mathrm{ST}(\pm)},
    I_y^{\mathrm{ST}(\pm)}
    \big]
    =i I_z^{\mathrm{ST}(\pm)}.
\end{equation}

For the symmetries in table~\ref{tab:Symmetries}, the first-order average Hamiltonian in equation~\ref{eq:H1CS} may be written as follows:
\begin{equation}
 \overline{H}^{(1)}_{\rm CS} 
 =
 \wSTnut\big(
 I_x^{\mathrm{ST}(+)}
    \cos\phi_\mathrm{ST}
+
 I_y^{\mathrm{ST}(+)}
    \sin\phi_\mathrm{ST}
 \big).
\end{equation}

Assume that the density operator of the spin ensemble is prepared with a population difference between the lower triplet state and the singlet state. This arises, for example, if the system is in thermal equilibrium in a strong magnetic field. This state corresponds to a density operator term of the form:
\begin{equation}
    \rho(0) \sim I_z^{\mathrm{ST}(+)}
\end{equation}
omitting numerical factors and orthogonal operators. Suppose that an integer number $p$ of complete \RNnnu sequences is applied, with symmetry numbers selected from table~\ref{tab:Symmetries}. The excitation interval is given by $\tau=pT$, where $T=N\tau_R$ is the duration of a complete \RNnnu sequence. From the cyclic commutation relationships in equation~\ref{eq:STCyComm}, the density operator at the end of the sequence is given by
\begin{align}
\label{eq:rhotau}
\rho(\tau) 
    &\simeq 
I_z^{\mathrm{ST}(+)}\cos(\wSTnut\tau)
\nonumber\\
&- I_x^{\mathrm{ST}(+)}\sin(\wSTnut\tau)
\cos(\phi_\mathrm{ST})
\nonumber\\
&+ I_y^{\mathrm{ST}(+)}\sin(\wSTnut\tau)
\sin(\phi_\mathrm{ST}).
\end{align}
This suggests the following phenomena:
\begin{enumerate}
  %. . . . . 
 \item 
 \emph{Excitation of Singlet-Triplet Coherence}.
 If the interval $\tau$ is chosen such that $\wSTnut\tau$ is approximately an odd multiple of $\pi/2$, the resulting density operator contains terms proportional to the transverse operators $I_x^{\mathrm{ST}(+)}$ and $I_y^{\mathrm{ST}(+)}$, indicating the excitation of singlet-triplet coherence~\cite{sheberstov_excitation_2019}. 
 In practice, the evolution time $\tau^{*}$ is restricted to integer multiples of the basic element duration $\tau_R$. In the absence of dissipative effects, the excitation of a singlet-triplet coherence is optimized by completing the following number of R-elements:
\begin{equation}\label{eq:nstarSTexc}
\begin{aligned}
n^{*}
&\simeq{\rm round}
(\pi/(4\wSTnut\tau_{\rm R}))
\nonumber\\
&\quad\quad\quad
\mbox{(ST coherence excitation)}
\\
\end{aligned}
\end{equation}
 %. . . . . 
 \item \emph{Generation of Singlet Order.}
If the interval $\tau$ is chosen such that $\wSTnut\tau$ is approximately an even multiple of $\pi/2$, the term $I_z^{\mathrm{ST}(+)}$ is inverted in sign. This indicates that the populations of the singlet state and the outer triplet state are swapped. This leads to the generation of singlet order, which is a long-lived  difference in population between the singlet state and the triplet manifold~\citeLLS. 
In the absence of relaxation, the conversion of magnetization into singlet-order is optimised by completing the following number of R-elements:
\begin{equation}\label{eq:nstarSOgen}
\begin{aligned}
n^{*}\simeq{\rm round}(\pi/(2\wSTnut\tau_{\rm R}))
\quad
\mbox{(SO generation)}
\end{aligned}
\end{equation}
%. . . . . 
\end{enumerate}
It follows that the application of a \RNnnu sequence to a near-equivalent 2-spin-1/2 system in thermal equilibrium leads either to the excitation of singlet-triplet coherences, or to the generation of singlet order, depending on the number of R-elements that are applied. Experimental demonstrations of both effects are given below. 

There are technical complications if the number of applied R-elements does not correspond to an integer number of complete \RNnnu sequences. In such cases the operators $U_J$ and $U_\mathrm{rf}$ in equation~\ref{eq:U} lead to additional transformations. If the total number of completed R-elements is \emph{even}, the main consequence is an additional phase shift of excited coherences, which is often of little consequence. If the number of applied R-elements is \emph{odd}, on the other hand, then the propagator $U_\mathrm{rf}$ swaps the $\ket{T_{+1}}$ and $\ket{T_{-1}}$ states, exchanging the $I_z^{\mathrm{ST}(\pm)}$ operators.

%\MHLToHere

%--------------------------
%\input{Figures/Rseq-standard}
\subsection{Implementation}
\subsubsection{Standard Implementation}
The standard implementation of a \RNnnu sequence is sketched in figure~\ref{fig:Rseqstandard} and described by equation~\ref{eq:RNnnustand}.

There is great freedom in the choice of the basic element $\mathcal{R}^{0}$ upon which the sequence is constructed. In this paper we concentrate on the implementation shown in figure~\ref{fig:Rseqstandard}, in which the basic element is a three-component composite pulse~\cite{levitt_nmr_1979}, with two $\tau$ delays inserted between the pulses:
\begin{equation}\label{eq:R0}
    \mathcal{R}^{0} = 
      \left(  90_{90} -\tau- 180_{0} -\tau- 90_{90}
      \right)
\end{equation}
where degrees are used here for the flip angles and the phases. This composite pulse generates an overall rotation by $\pi$ around the rotating-frame x-axis~\cite{levitt_compensation_1981}, and hence is an eligible basic element $\mathcal{R}^0$ for the construction of a \RNnnu sequence. 

The scaling factor $\kappa_{1111}$, and hence the nutation frequency of the singlet-triplet transition, depends on the choice of basic element. In the case of the basic element in equation~\ref{eq:R0}, the scaling factor is readily calculated in the limit of ``$\delta$-function" pulses, i.e. strong rf pulses with negligible duration. The scaling factors $\kappa_{1 \pm1 1 \pm1}$ are given for general $N$, $n$ and $\nu$ by
\begin{equation}\label{eq:kappa}
\kappa_{1 \pm1 1 \pm1} =
    \ 2^{1/2}\frac{N}{n\pi}
    (-1)^{
       (N \pm (n-\nu))/(2N)}
        \sin^2 (n\pi /2N ).
\end{equation}
Scaling factors for a set of \RNnnu symmetries appropriate for singlet-triplet excitation are given in table~\ref{tab:Symmetries}. Scaling factors with the largest magnitude are offered by sequences with the symmetries \Rsymm431,  \Rsymm851, \Rsymm873, and \Rsymm{10}72.

Since the scaling factors in equation~\ref{eq:kappa} are real, the effective nutation axis of the singlet-triplet transition has a phase angle of zero, $\phi_\mathrm{ST}=0$. This result applies to the basic-R element in equation~\ref{eq:R0}, in the $\delta$-function pulse limit.   

The implementation of a \RNnnu sequence by the procedure in figure~\ref{fig:Rseqstandard} provides selective excitation of the transition between the singlet state of a near-equivalent spin-1/2 pair and one of the outer triplet states. However, the sequence performance is not robust with respect to rf field errors. It is readily shown that a deviation of the rf field from its nominal value induces a net rotation around the z-axis which accumulates as the sequence proceeds. This causes a degradation in performance in the case of radiofrequency inhomogeneity or instability.  

%------------------------
\begin{figure}[tb]
\includegraphics[width=0.5\textwidth]{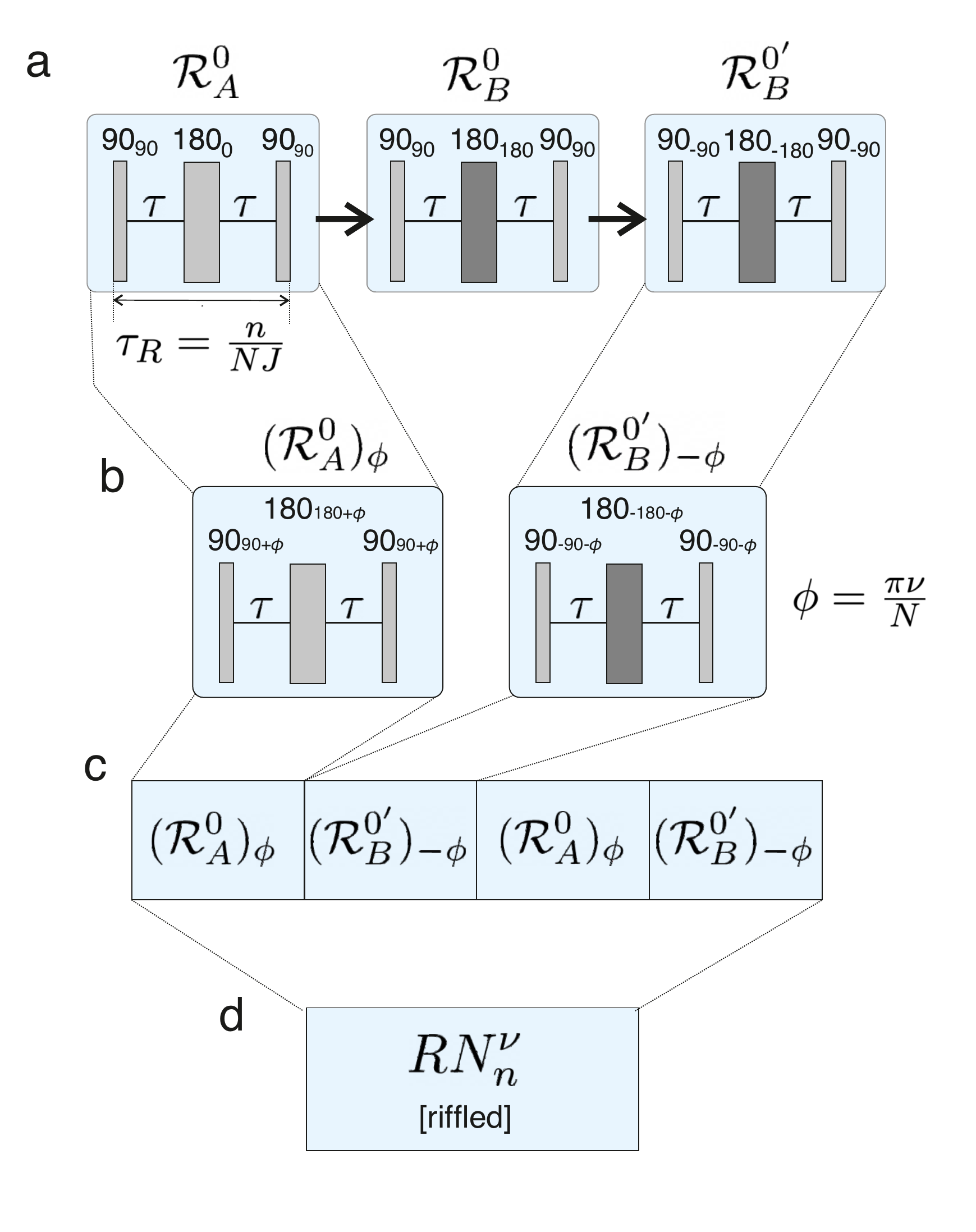}
\caption{
\label{fig:RseqPP}
The construction of a riffled \RNnnu sequence for singlet-triplet conversion. (a) Two basic $R$-elements are used; The elements $\mathcal{R}_A^0$ and $\mathcal{R}_B^0$ have identical properties under suitable approximations, but have opposite responses to pulse imperfections. In the current case, $\mathcal{R}_A^0$ is given by the composite pulse $\mathrm{90_{90} 180_0 90_{90}}$ with delays $\tau$ between the pulses, such that its overall duration is $\tau_R=n/(NJ)$. The element $\mathcal{R}_B^0$ is identical but with a $180\deg$ phase shift of the central pulse (dark shade). The conjugate sequence $\mathcal{R}_B^{0'}$ is generated from $\mathcal{R}_B^{0}$ by a change in sign of all phases. (b) The sequence $\mathcal{R}_A^0$ is given a phase shift of $+\phi$, while the sequence $\mathcal{R}_B^{0'}$ is given a phase shift of $-\phi$, where $\phi=\pi\nu/N$. (c) The pair of sequences $(\mathcal{R}_A^0)_\phi$ and $(\mathcal{R}_B^{0'})_{-\phi}$ is repeated $N/2$ times, to give a riffled \RNnnu sequence (d). PulsePol is an example of a riffled \RNnnu sequence (see text). 
}
\end{figure}
\subsubsection{Riffled Implementation}

%\blue{
In magic-angle-spinning NMR, error compensation is often achieved by the use of supercycles, i.e. repetition of the entire sequence with variations in the phase shifts, or in some cases, cyclic permutations of the pulse sequence elements~\cite{brinkmann_synchronous_2000,brinkmann_homonuclear_2002,kristiansen_robust_2004,brouwer_symmetry-based_2005,kristiansen_theory_2006}.  PulsePol achieves very effective compensation for rf pulse errors by a much simpler method, namely a phase shift of just one pulse by $180\deg$. This simple modification may be interpreted as a modified procedure for constructing sequences with \RNnnu symmetry, but with built-in error compensation. 
%}

%\blue{
Consider two different basic elements, denoted here $\mathcal{R}^0_A$ and $\mathcal{R}^0_B$, as shown in figure~\ref{fig:RseqPP}a. In the depicted case, the two basic elements differ only in that the central 180\deg pulse is shifted in phase by 180\deg: 
\begin{align}\label{eq:R0AB}
    \mathcal{R}^{0}_A &= 
      \left(  90_{90} -\tau- 180_{0} -\tau- 90_{90}
      \right)
\nonumber\\
     \mathcal{R}^{0}_B &= 
      \left(  90_{90} -\tau- 180_{180} -\tau- 90_{90}
      \right)
\end{align}
Under ideal conditions, both of these basic elements provide a net rotation by an odd multiple of $\pi$ about the rotating-frame x-axis, and hence are eligible starting points for the \RNnnu construction procedure. Furthermore, \emph{in the $\delta$-function pulse limit}, the Euler angle trajectories generated by these sequences are identical. This implies that, in the case of ideal, infinitely short pulses, the elements $\mathcal{R}^{0}_A$ and $\mathcal{R}^{0}_B$ are completely interchangeable. The modified \RNnnu construction procedure sketched in figure~\ref{fig:RseqPP} exploits this freedom by alternating the phase shifted ``A" basic element $(\mathcal{R}^{0}_A)_{+\pi\nu/N}$ with the phase-shifted conjugate ``B" element $(\mathcal{R}^{0'}_B)_{-\pi\nu/N}$. 
%}

The alternation of two different basic elements, as shown in figure~\ref{fig:RseqPP}, resembles the ``riffling" technique for shuffling a pack of cards, in which the pack is divided into two piles, and the corners of the two piles are flicked up and released so that the cards intermingle. The procedure in figure~\ref{fig:RseqPP} therefore leads to a \emph{riffled \RNnnu sequence}.

%\blue{
Under ideal conditions, and for pulses of infinitesimal duration, the ``standard" and ``riffled" construction procedures have identical performance. However, an important difference arises in the presence of rf field amplitude errors. The errors accumulate in the ``standard" procedure, but cancel out in the ``riffled" procedure. Hence the procedure shown in figure~\ref{fig:RseqPP} achieves more robust performance with respect to rf field errors than the standard procedure of figure~\ref{fig:Rseqstandard}. However, it should be emphasised that this form of error compensation does not apply to all basic R-elements, and that even in the current case, strict \RNnnu symmetry is only maintained in the limit of $\delta$-function pulses. Nevertheless, within these caveats and restrictions, this error-compensation procedure is powerful and useful. As discussed below, error-compensation by riffling is responsible for the robust performance of PulsePol. 

%\blue{
To see how a PulsePol sequence~\cite{schwartz_robust_2018,tratzmiller_pulsed_2021,tratzmiller_parallel_2021} arises from the riffled \RNnnu construction procedure, start with the pair of basic R-elements given in equation~\ref{eq:R0AB}. Consider the symmetry \Rsymm431, which is appropriate for transition-selective singlet-triplet excitation, as shown in table~\ref{tab:Symmetries}. This symmetry implies that each R-element has duration $\tau_R = (3/4)J^{-1}$, and hence that the delays between the pulses are given by $\tau=\tau_R/2=(3/8)J^{-1}$, in the $\delta$-function pulse limit. 
%}

%\blue{
The phase shifts $\pm\pi\nu/N$ are equal to $\pm45\deg$ in the case of \Rsymm431 symmetry. Hence the pair of phase-shifted elements is given by
\begin{align}
    (\mathcal{R}^{0}_A)_{+45} &= 
      \left(  90_{135} -\tau- 180_{45} -\tau- 90_{135}
      \right)
\nonumber\\
     (\mathcal{R}^{0'}_B)_{-45} &= 
      \left(  90_{-135} -\tau- 180_{-225} -\tau- 90_{-135}
      \right)
\end{align}
This pair of elements may be concatenated, and the pair of elements repeated, to complete the riffled implementation of \Rsymm431:
\begin{equation}\label{eq:R431riffled}
    \Rsymm431 \mathrm{\,[riffled]}=
  (\mathcal{R}^{0}_A)_{+45}
  (\mathcal{R}^{0'}_B)_{-45}
  (\mathcal{R}^{0}_A)_{+45}
  (\mathcal{R}^{0'}_B)_{-45}
\end{equation}
%}

%\blue{
If the riffled \Rsymm431 sequence is given a $-45\deg$ phase shift, we get:
\begin{align}\label{eq:PP}
   & \Big[
    (\mathcal{R}^{0}_A)_{+45}
    (\mathcal{R}^{0'}_B)_{-45}
    \Big]_{-45}
= 
    (\mathcal{R}^{0}_A)_{0}
    (\mathcal{R}^{0'}_B)_{-90}
\nonumber\\ 
= &\ %
    \left(
    90_{90} -\tau- 180_{0} -\tau- 90_{90}
    \cdot
    90_{0} -\tau- 180_{90} -\tau- 90_{0}
    \right)
\nonumber\\ 
\end{align}
which is PulsePol~\cite{schwartz_robust_2018,tratzmiller_pulsed_2021,tratzmiller_parallel_2021}.
The $-45\deg$ phase shift is of no consequence for the interconversion of singlet order and magnetization.
%
%}

%\blue{
The riffled construction procedure may be deployed for the other symmetries in table~\ref{tab:Symmetries}. For example, the riffled implementation of \Rsymm873, using the basic elements in equation~\ref{eq:R0AB}, is as follows:
\begin{align}\label{eq:R873PP}
\Rsymm873\mbox{\ [riffled]}
=&\left[
(\mathcal{R}^{0}_A)_{+67.5}
    (\mathcal{R}^{0'}_B)_{-67.5}
\right]^{4}
\nonumber\\
=&\big[
90_{157.5} -\tau- 180_{67.5} -\tau- 90_{157.5}
\ \cdot
\nonumber\\
& 90_{-157.5} -\tau- 180_{112.5} -\tau- 90_{-157.5}
\big]^4
\end{align}
where the superscript indicates 4 repetitions and the interpulse delays are given by $\tau = \tau_R/2 = (7/8)J^{-1}$, in the $\delta$-function pulse limit. Some sequences of this type have been proposed in the form of ``generalised PulsePol sequences"~\cite{tratzmiller_parallel_2021,tratzmiller_pulsed_2021}. 
%}

%\blue{
The performance of these sequences may be made even more robust by using composite pulses for the 90\deg or 180\deg pulse sequence elements~\citeCompPulse. 
Some examples are demonstrated below. 

%%---Figure----
%%---
\section{\label{sec:Procedure}Experimental}
%--------------------
\subsection{Sample}

%================================================
%%%  start_table:molecules  %%%
%================================================
\begingroup
\setlength{\tabcolsep}{1pt}
\renewcommand{\arraystretch}{1.2}
\begin{table}
  \caption{Chemical structure of \Ctwo-DAND (1,2,3,4,5,6,8-heptakis(methoxy-\emph{d}$_3$)-7-((propan-2-yl-\emph{d}$_7$)oxy)naphthalene-4a,8a-[\Ctwo]) with its relevant NMR parameters in a magnetic field of $9.39$~T. The singlet-triplet mixing angle is defined as $\theta_{\rm ST}=\tan^{-1}{(\omega_{\Delta}/{2\pi J})}$~ \cite{bengs_generalised_2020}. %\MHLnote{give the definition and a reference.}
  }
    \label{tab:CS_molecules}
\centering
\begin{tabular*}{0.95\columnwidth}{ll}
\Xhline{2.5\arrayrulewidth}
\begin{minipage}[t]{0.95\columnwidth}
\raggedright
\includegraphics[trim={1cm 3cm 1cm 3cm},clip,width=\columnwidth]{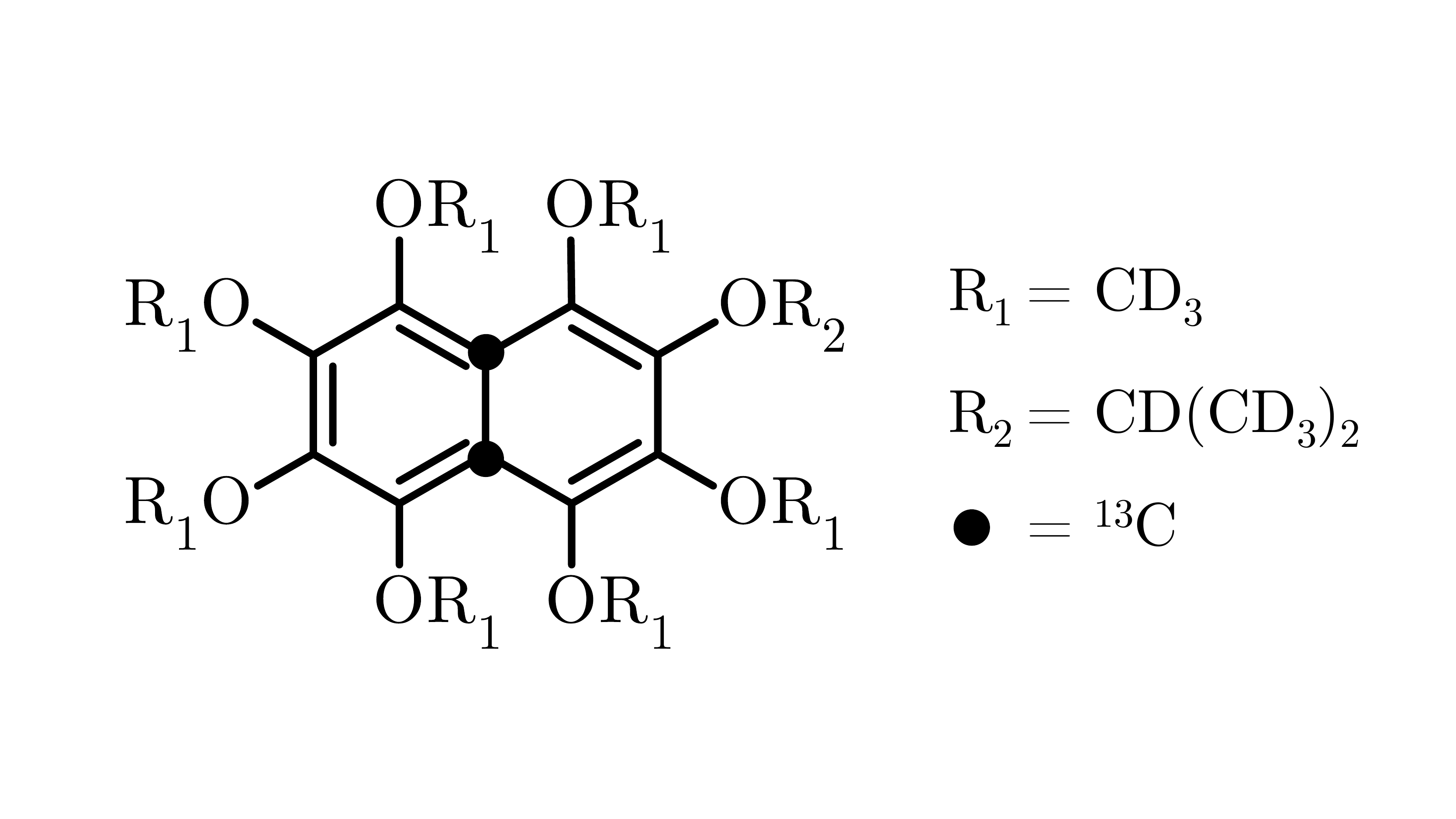}
\end{minipage}
\\
\hline
$J_{\rm CC}/{\rm Hz}$ & \hspace{-100 pt}$54.39\pm0.10$
\\
\hline
$\Delta \delta / {\rm ppb} $ & \hspace{-100 pt}$75.0 \pm 2.0 $
\\
\hline
$\omega_{\Delta}/(2\pi) /{\rm Hz}$ {[}{\rm @9.4 T}{]} & \hspace{-100 pt}$7.50\pm0.20$
\\
\hline
$\theta_{\rm ST}/{}^{\circ}$ & \hspace{-100 pt}$7.85\pm0.22$
\\
\hline
\end{tabular*}
\end{table}
\endgroup
%================================================
%%%    end_figure:molecules    %%%
%================================================

%\blue{
Experiments were performed on a solution of a \Ctwo-labelled deutero-alkoxy naphthalene derivative (\Ctwo-DAND), whose molecular structure with its relevant NMR parameters is shown in table~\ref{tab:CS_molecules}. Further details of the synthesis of (\Ctwo-DAND) are given in the reference by Hill-Cousins et al~\cite{hill-cousins_synthesis_2015}. This compound exhibits a very long \Ctwo singlet lifetime in low magnetic field~\cite{stevanato_nuclear_2015}. The current experiments were performed on 30 mM of \Ctwo-DAND dissolved in 500 \uL \disopropanol. The two \Cth sites have a J-coupling of 54.39$\pm$0.10~Hz and a chemical shift difference of 7.50$\pm0.2$~Hz in a magnetic field of $9.39$~T. The solution was doped with 3 mM of the paramagnetic agent (2,2,6,6-tetramethylpiperidin-1-yl)oxyl (TEMPO) in order to decrease the $\Tone$ relaxation time, allowing faster repetition of the experiments, and was contained in a standard Wilmad 5 mM sample tube.

\subsection{NMR Equipment}
All spectra were acquired at a magnetic field of 9.39~T. A 10~mm NMR probe was used,
with the radiofrequency amplitude adjusted to give a nutation frequency of $\wnut/(2\pi)\simeq$12.5~kHz, corresponding to a 90\deg pulse duration of $20 \,\mu\mathrm{s}$.

\begin{figure}[th]
\includegraphics[width=0.5\textwidth]{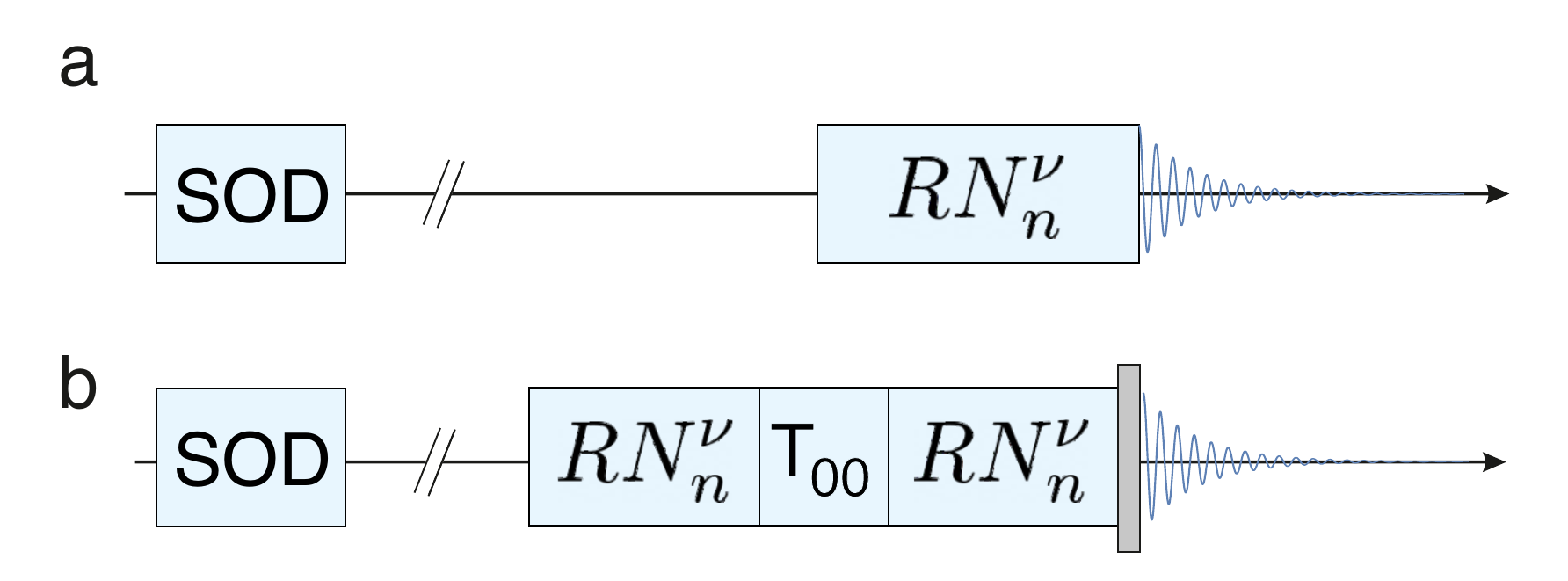}
\caption{
\label{fig:PulseSequences}
High-field NMR pulse sequences used in this work. (a) 
After a singlet-order destruction sequence (SOD)~\cite{rodin_SOD_2019} and a waiting interval to establish thermal equilibrium, a \RNnnu sequence is applied to thermal equilibrium magnetization, exciting coherences between the singlet state and one of the outer triplet states. (b) Procedure for estimating singlet order generation. A \RNnnu sequence is applied to generate singlet order, followed by a \Tzz singlet-order-filtering sequence~\cite{tayler_accessing_2013,tayler_theory_2012}, and a second \RNnnu sequence to regenerate $z$-magnetization. The NMR signal is induced by applying a composite $90\deg$ pulse (grey rectangle). 
%The \Tzz sequence destroys any signals that do not pass through rank-0 singlet order at the junction of the two \RNnnu sequences. }
%\MHLnote{caption needs changing to reflect revised figure.}
}
\end{figure}
\subsection{Pulse Sequences}
\subsubsection{Singlet-Triplet Excitation}
The excitation of coherences between the singlet state and the outer triplet states of \Ctwo-DAND was demonstrated using the pulse sequence in figure~\ref{fig:PulseSequences}a.
On each transient, a singlet destruction block~\cite{rodin_SOD_2019} is applied followed by a waiting time of $\sim5\Tone$ to establish thermal equilibrium. This ensures an initial condition free from interference by residual long-lived singlet order left over from the previous transient. 
After thermal equilibration in the magnetic field, a \RNnnu symmetry-based singlet-triplet excitation sequence of duration $\tauexc$ is applied and the NMR signal detected immediately afterwards. Fourier transformation of the signal generates the \Cth NMR spectrum. 

 \begin{table}[b]
 \renewcommand{\arraystretch}{1.5}
  \begin{tabular*}{0.4\textwidth}{@{\extracolsep{\fill}}lll}
\hline
    $\omega_{nut}/(2\pi)$ & $12.5$ kHz
\\
    $\tau_{90}$ & 20 $\mu$s
\\
\hline
    $\tau_R$    & 13800 ${\, \mu \mathrm{s}}$
\\
    $\tau$    & 6860 ${\,\mu \mathrm{s}}$
\\
    $n_R^\mathrm{exc}$       &  4
\\
    $\tau_\mathrm{exc}$     & 55.2 ms
\\ \hline
  \end{tabular*}
 \caption{
 \label{tab:OuterSTexcPars}
 Experimental parameters for the $\mathrm{R4_3^{\pm1}}$ sequences used to obtain the results in figure~\ref{fig:OuterSTexc}(c,d). The parameters have the following meaning:
 $\wnut$ is the radiofrequency pulse amplitude, expressed as a nutation frequency; $\tau_{90}$ is the duration of a $90\deg$ pulse; $\tau_R$ is the duration of a single R-element; $\tau$ is the interval between pulses within each R-element (see figure~\ref{fig:Rseqstandard}); $n_R^\mathrm{exc}$ is the number of R-elements in the excitation sequence; $\tau_\mathrm{exc}$ is the duration of the excitation sequence. 
% \MHLnote{The table gave $\tau_{90}=40{\,\mu s}$ which is inconsistent with the specified nutation frequency. So which is correct?}
}
\end{table}
\begin{figure}[tb]
\centering
\includegraphics[trim={0 0 0 0},clip,width=0.9\columnwidth]{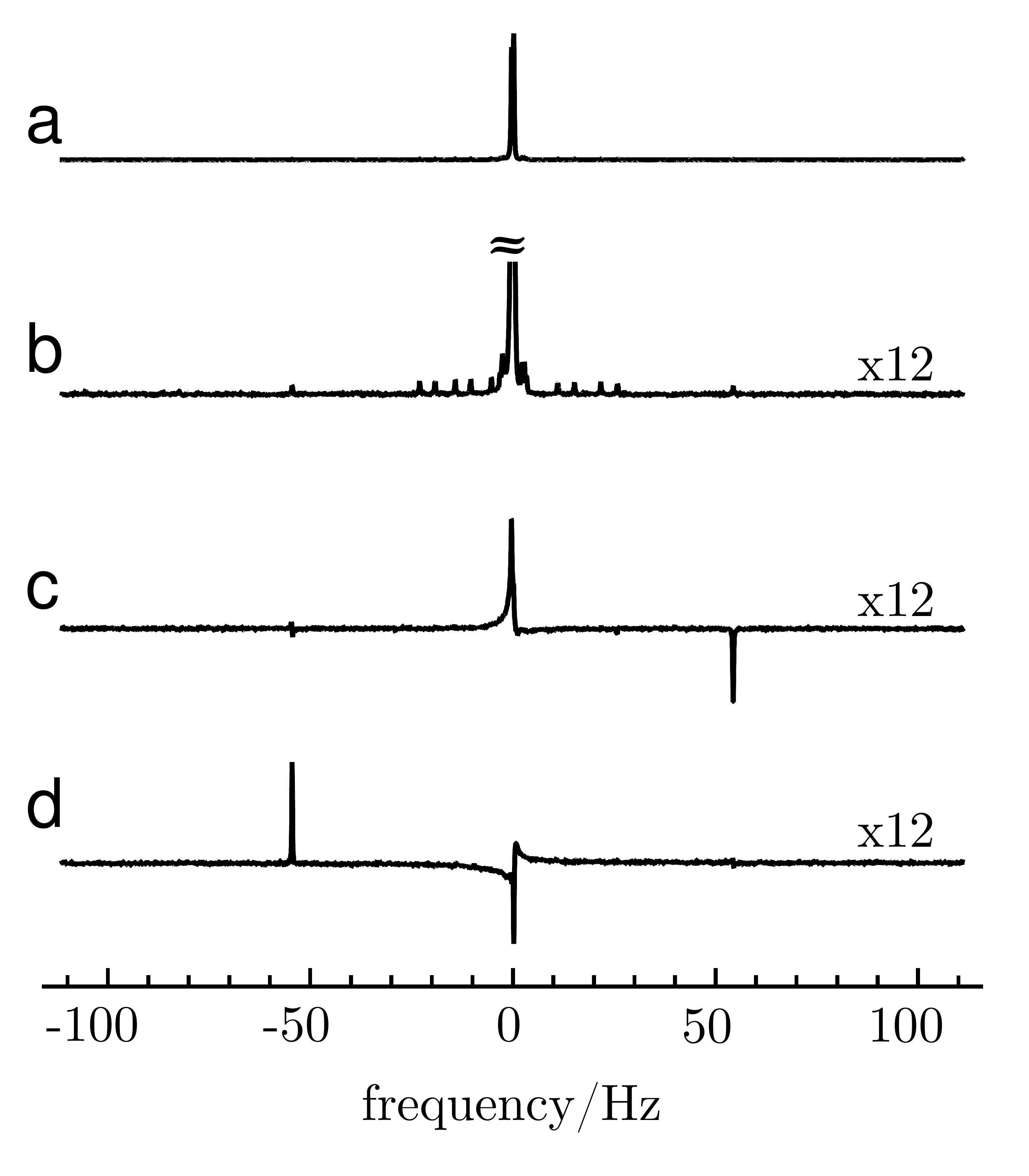}
\caption{\label{fig:OuterSTexc} 
Enhanced singlet-triplet coherent excitation. 
(a) Conventional \Cth spectrum of \Ctwo-DAND using a single 90\deg pulse for excitation, showing strong signals from the triplet-triplet coherences;
%\MHLnote{expand the vertical scale of this! ffs!}
(b) Vertical expansion (by a factor of 12) %\MHLnote{decrease this factor when (a) is expanded!} 
of the conventional \Cth spectrum. Additional signals are visible from minority isotopomers, with the outer peaks barely visible. The strong central peak is truncated. (c) Spectrum obtained by applying four elements of a riffled \Rsymm431 sequence, showing a strongly enhanced outer peak. The  construction procedure in figure~\ref{fig:RseqPP} was used, starting from the basic elements in equation~\ref{eq:R0AB}. (d) Spectrum obtained by applying four elements of a \Rsymm43{-1} sequence, showing the enhancement of the other outer peak. All spectra were obtained with a total of 256 transients and the same processing parameters. No line broadening is applied.
%\MHLnote{why is (a) so small? Once again, there is an irritating poor choice of format, obvious to anyone, and for no discernible reason. Decrease the vertical scale in (a) by a factor of approx.4, and change the annotation of the expansion factors of the other spectra accordingly.}
}
\end{figure}

\subsubsection{Singlet Order Generation}
The generation of singlet order is assessed by the pulse sequence scheme in figure~\ref{fig:PulseSequences}b. After destruction of residual singlet order and thermal equilibration, a M2S or \RNnnu sequence of duration $\tauexc$ is applied to generate singlet order. This is followed by a \Tzz singlet filter sequence~\cite{tayler_singlet_2011}. This consists of a sequence of rf pulses and pulsed field gradients that dephase all signal components not associated with nuclear singlet order. The singlet order is reconverted to z-magnetization by a second \RNnnu sequence of equal duration to the first, or by a S2M sequence (time-reverse of the M2S sequence)~\cite{pileio_storage_2010,tayler_singlet_2011}. The recovered z-magnetization is converted to transverse magnetization by a composite 90\deg pulse and the NMR signal detected in the following interval. The signal amplitude serves as a measure of the singlet order generated by the excitation sequence, and the efficiency of recovering magnetization from the singlet order.  The maximum theoretical efficiency for passing magnetization through singlet order is $2/3$~\cite{levitt_symmetry_2016}.

The \RNnnu sequences may be constructed by either the standard or the riffled procedures. M2S and S2M sequences may be substituted for the first and last \RNnnu sequences, respectively. The $90\deg$ readout pulse in figure~\ref{fig:PulseSequences}b was implemented as a symmetrized BB1 composite pulse~\cite{wimperis_broadband_1994,cummins_tackling_2003}. Details of the composite pulse, the SOD sequence, and the \Tzz pulse sequence modules are given in the Supporting Information.

%----------------------
\begin{figure}[tb]
\centering
\includegraphics[trim={0.75cm 0 0.75cm 0 },clip,width=\columnwidth]{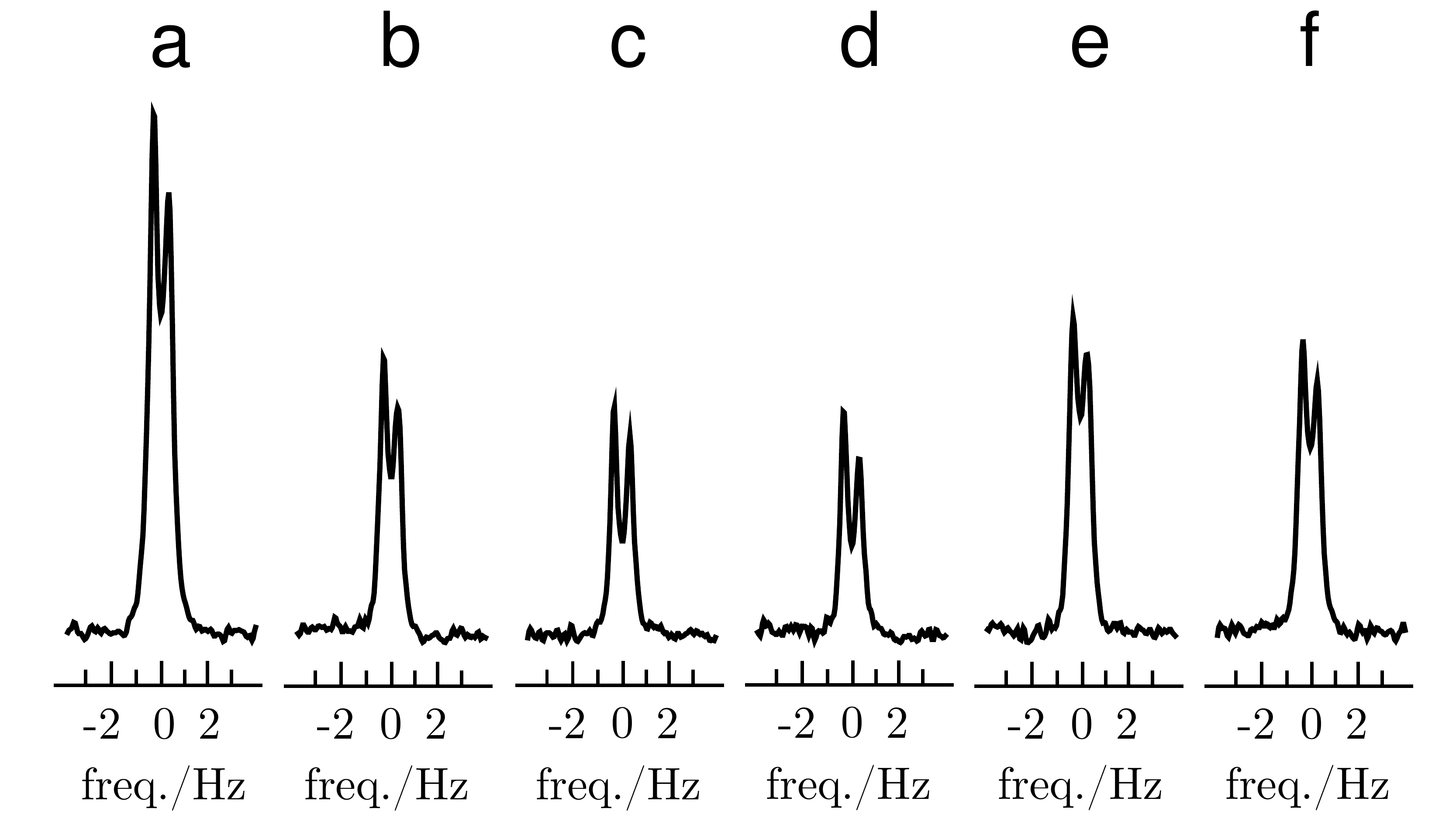}
\caption{\label{fig:SingletFilteredSpectra} \Cth spectra obtained after (a) a single $90\deg$ pulse, or (b-f) after filtering the \Cth NMR signal through singlet order, using the scheme in figure~\ref{fig:PulseSequences}b. (a) Standard \Cth spectrum obtained with a single $90\deg$ pulse. (b) Singlet-filtered spectrum obtained with M2S for singlet order excitation and S2M for reconversion to magnetization. (c) Singlet-filtered spectrum obtained with a pair of \Rsymm431 sequences.
(d) Singlet-filtered spectrum obtained with a pair of \Rsymm873 sequences. Both (c) and (d) use the standard implementation of \RNnnu sequences, as in  figure~\ref{fig:Rseqstandard}, using the basic element in equation~\ref{eq:R0}). (e) Singlet-filtered spectrum obtained with a pair of riffled \Rsymm431 sequences. (f) Singlet-filtered spectrum obtained with a pair of riffled \Rsymm873 sequences. Both (e) and (f) use the riffled implementation of \RNnnu sequences, as in figure~\ref{fig:RseqPP}, using the basic elements in equation~\ref{eq:R0AB}. All pulse sequence parameters are given in the Supporting Information.
}
\end{figure}
\section{\label{sec:results}Results}
\subsection{Transition-selective singlet-triplet excitation}

%\blue{
In systems of near-equivalent spin-1/2 pairs, the chemical shift difference induces a slight mixing of the singlet state $\ket{S_0}$ with the central triplet state $\ket{T_0}$. This effect lends signal intensity to the single-quantum coherences between the singlet state and the outer triplet states $\ket{T_{\pm1}}$, which generate the outer lines of the AB quartet. These peaks are feeble for two independent reasons: (i) the coupling of the singlet-triplet coherences to observable transverse magnetization is weak in the near-equivalence limit, and (ii) the singlet-triplet coherences are excited only weakly by conventional single-pulse excitation. The first of these factors is an  intrinsic property of a singlet-triplet coherence. The second factor, on the other hand, may be overcome by using a suitable excitation sequence to generate the desired coherence with full amplitude. Many such schemes have been devised~\cite{sheberstov_excitation_2019}. This effect is useful since the frequencies of these peaks provide an accurate estimate of the internuclear J-coupling, which can be difficult to estimate in the near-equivalence regime. 

%}

Figure~\ref{fig:OuterSTexc}a shows the \Cth NMR spectrum of the \Ctwo-DAND solution. The strong central doublet is due to the two triplet-triplet coherences. The outer peaks of the AB quartet, which correspond to the weakly allowed singlet-triplet coherences, are barely visible in the spectrum, even after vertical expansion (figure~\ref{fig:OuterSTexc}b). 

Greatly enhanced excitation of the outer AB peaks is achieved by the pulse sequence in figure~\ref{fig:PulseSequences}a, using an excitation sequence of symmetry \Rsymm431 constructed by the riffled procedure (figure~\ref{fig:RseqPP}), and with the number of R-elements satisfying equation~\ref{eq:nstarSTexc}. The strong enhancement of the outer AB peaks, relative to the spectrum induced by a single 90\deg pulse, is self-evident in figure~\ref{fig:OuterSTexc}c. Note that changing the sign of the symmetry number $\nu$ switches the excitation to the opposite singlet-triplet transition (figure~\ref{fig:OuterSTexc}d). The experimental pulse sequence parameters are given in table~\ref{tab:OuterSTexcPars}.

%=================
\begin{figure*}[tbh!]
\centering
\includegraphics[trim={0 0 0 0},clip,width=0.95\textwidth]{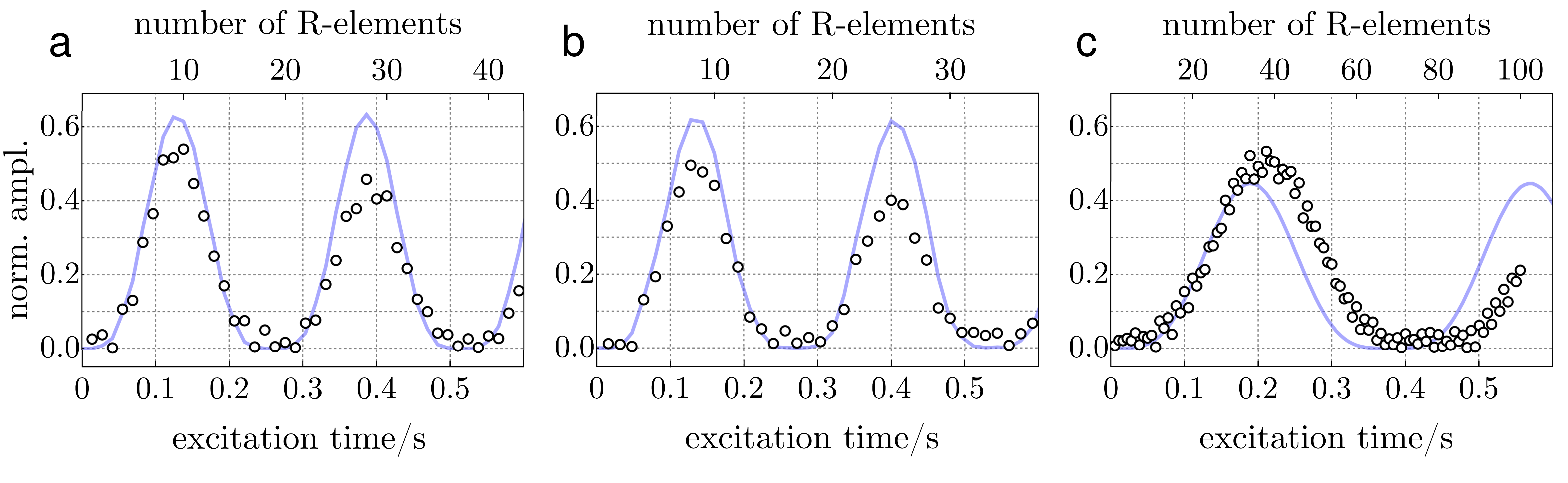}
\caption{\label{fig:SignalTrajectories} 
Experimental $^{13}{\rm C}$ signal amplitudes (white dots) for the protocol in figure~\ref{fig:PulseSequences}b, using riffled \RNnnu sequences for both the excitation and reconversion of singlet order. The following symmetries were used: (a) \Rsymm431, (b) \Rsymm873 and (c) \Rsymm{10}32. 
The number $n_R$ of R-elements in the  \RNnnu sequences for singlet excitation and reconversion are varied simultaneously (top horizontal axis). The corresponding total duration of each sequence is shown on the lower horizontal axis. All sequences were implemented by the riffled procedure in figure~\ref{fig:RseqPP}, using the basic elements in equation~\ref{eq:R0AB}. The signal amplitudes are normalized relative to that generated by a single 90\deg pulse. Light blue trajectories show numerical simulations (excluding relaxation) with the pulse sequence parameters given in the SI.
}
\end{figure*}

\subsection{Magnetization-to-singlet conversion}

The experimental performance of some magnetization-to-singlet conversion schemes was tested on a TEMPO-doped solution of \Ctwo-DAND using the pulse sequence protocol in figure~\ref{fig:PulseSequences}b. A selection of singlet-filtered NMR spectra is shown in figure~\ref{fig:SingletFilteredSpectra}(b-f). In all cases the pulse sequence parameters were optimised for the best performance. The optimised parameters are given in the Supporting Information. 

Figure~\ref{fig:SingletFilteredSpectra}a shows the unfiltered \Cth NMR spectrum of \Ctwo-DAND. Figure~\ref{fig:SingletFilteredSpectra}b shows the spectrum obtained by applying a M2S sequence to generate singlet order, suppressing other spin order terms, and regenerating magnetization from singlet order by applying a S2M sequence. Approximately 50\% of the spin order is lost by this procedure, as may be seen by comparing the spectra in figure~\ref{fig:SingletFilteredSpectra}a and b. The theoretical limit on passing magnetization through singlet order is $2/3\simeq67\%$.

The results obtained by using \RNnnu sequences with different sets of symmetry numbers are shown in figure~\ref{fig:SingletFilteredSpectra}c and d. The standard \RNnnu construction procedure in figure~\ref{fig:Rseqstandard} was used. The number of R-elements was selected according to equation~\ref{eq:nstarSOgen}. The results are slightly inferior to the M2S sequence. Some of these spectra exhibit perturbed peak intensities. This is unexplained. 

Riffled \RNnnu sequences constructed by the procedure in figure~\ref{fig:RseqPP} display an improved performance, which is distinctly superior to M2S, as shown in figure~\ref{fig:SingletFilteredSpectra}e and f. The improvement is attributed to the increased robustness of the riffled procedure with respect to a range of experimental imperfections, as discussed further below. 

Note that the riffled \Rsymm431 sequence only differs from PulsePol~\cite{schwartz_robust_2018,tratzmiller_pulsed_2021,tratzmiller_parallel_2021} by an overall phase shift (equations~\ref{eq:R431riffled} and \ref{eq:PP}). The increased robustness of PulsePol with respect to M2S/S2M in the context of singlet/triplet conversion has been anticipated by the simulations of Tratzmiller~\cite{tratzmiller_pulsed_2021}.

The singlet order relaxation time $T_S$ is readily estimated by introducing a variable delay before the second \RNnnu sequence
in figure~\ref{fig:PulseSequences}b. Some results are shown in the Supporting Information. Although $T_S$ is found to be much greater than $T_1$, the value of $T_S$ is considerably shorter than that found in previous experiments~\cite{stevanato_nuclear_2015}. This is attributed to the TEMPO doping of the solution in the current case. 

\begin{figure}[tbh]
\centering
\includegraphics[trim={0 0 0 0},clip,width=0.95\columnwidth]{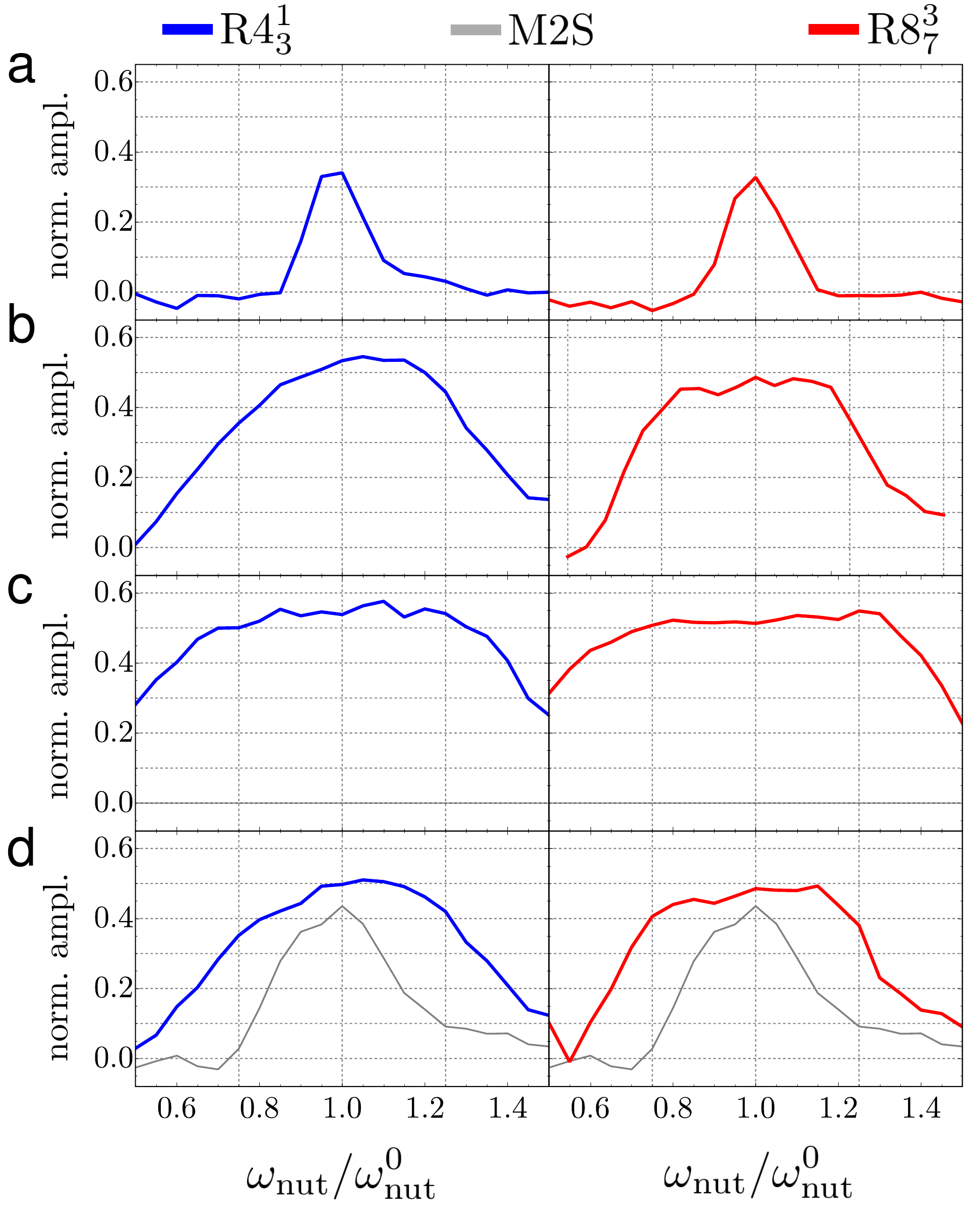}
\caption{\label{fig:RfFieldDependence}
Experimental \Cth signal amplitudes of \Ctwo-DAND solution, obtained by the protocol in figure~\ref{fig:PulseSequences}b, as a function of relative nutation frequency $\wnut/\wnut^0$, where $\wnut^0$ represents the nominal nutation frequency used for the calculation of pulse durations. The traces correspond to the experimental amplitudes for converting magnetization into singlet order and back again, normalized with respect to the signal generated by a single $90\deg$ pulse. Left column (blue): \Rsymm431 sequences; Right column (red): \Rsymm873 sequences. (a) Standard \RNnnu sequences using the basic element in equation~\ref{eq:R0}. (b) Riffled \RNnnu sequences using the basic elements in equation~\ref{eq:R0AB}. (c) Riffled \RNnnu sequences with all central $180_{0}$ pulses replaced by an ASBO-11 composite pulse~\cite{odedra_dual-compensated_2012}.  (d) Riffled \RNnnu sequences with all central $180_{0}$ pulses replaced by a $60_{180}180_{0}240_{180}420_{0}240_{180}180_{0}60_{180}$ composite pulse~\cite{shaka_symmetric_1987}. The grey lines in (d) show the experimental response of the M2S/S2M protocol. All experimental details are given in the SI. 
%\MHLnote{Replace $\Omega^{1}$ and $\Omega^{1}_{0}$ by $\wnut$ and $\wnut^0$.}
%\CBnote{Okay, no problem}
%\MHLnote{one curve has the wrong colour.}\CBnote{True, I missed that.}
}
\end{figure}
\begin{figure}[tbh]
\centering
\includegraphics[trim={0 0 0 0},clip,width=0.95\columnwidth]{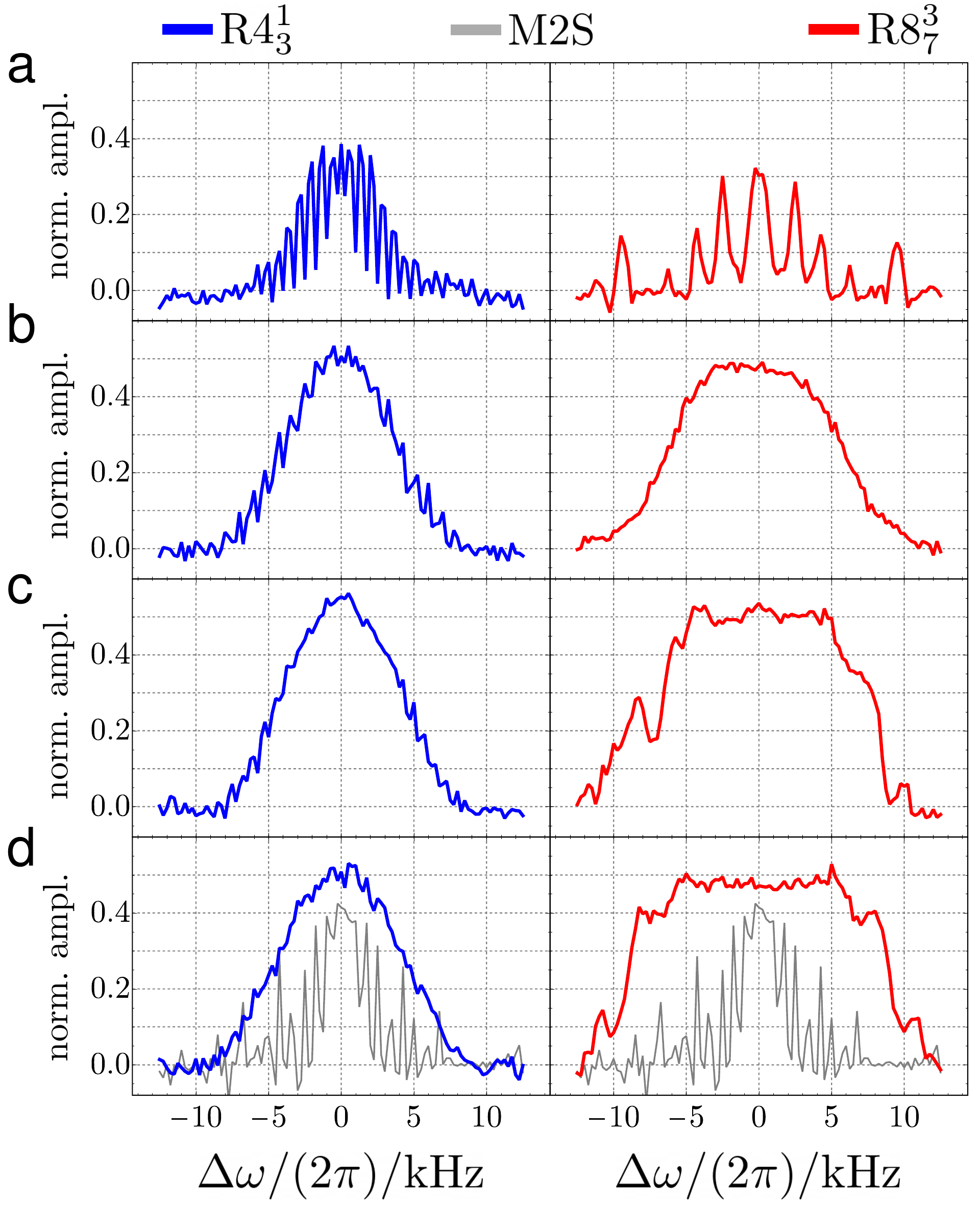}
\caption{
\label{fig:ResonanceOffsetDependence} Experimental \Cth signal amplitudes of \Ctwo-DAND solution, obtained by the protocol in figure~\ref{fig:PulseSequences}b, as a function of resonance offset $\Delta\omega$. The plotted points correspond to the amplitude for converting magnetization into singlet order and back again, normalized with respect to the signal generated by a single $90\deg$ pulse. Left column (blue): \Rsymm431 sequences; Right column (red): \Rsymm873 sequences. (a) Standard \RNnnu sequences using the basic element in equation~\ref{eq:R0}. (b) Riffled \RNnnu sequences using the basic elements in equation~\ref{eq:R0AB}. (c) Riffled \RNnnu sequences with all central $180_{0}$ pulses replaced by an ASBO-11 composite pulse~\cite{odedra_dual-compensated_2012}.  (d) Riffled \RNnnu sequences with all central $180_{0}$ pulses replaced by a $60_{180}180_{0}240_{180}420_{0}240_{180}180_{0}60_{180}$ composite pulse~\cite{shaka_symmetric_1987}. The grey lines in (d) show the experimental response of the M2S/S2M protocol. All experimental details are given in the SI. 
}
\end{figure}

Figure~\ref{fig:SignalTrajectories} shows the dependence of the singlet-filtered NMR signals on the number of R-elements $n_R$, used for both the excitation and reconversion sequence. The corresponding total sequence durations $\tau_\mathrm{exc}=\tau_\mathrm{recon}=n_R\tau_R = n_R(n/N)J^{-1}$ are also shown. Clear oscillations of the singlet order are observed, as predicted by equation~\ref{eq:rhotau}. The singlet order oscillations induced by \Rsymm873 are slightly slower than those for \Rsymm431, as expected from the theoretical scaling factors reported in table~\ref{tab:Symmetries}. 
The \Rsymm{10}32 sequence induces a relatively slow oscillation, corresponding to the small value of $\kappa_{1111}$ for this symmetry. In all cases, numerical simulations by \emph{SpinDynamica} software~\cite{bengs_spindynamica_2018} show qualitative agreement with the experimental results.

The improved robustness of the riffled implementation of \RNnnu with respect to rf amplitude variations is illustrated by the experimental results in figure~\ref{fig:RfFieldDependence}. These plots show the singlet-filtered signal amplitudes as a function of rf field amplitude, using the protocol in figure~\ref{fig:PulseSequences}b. Two different pulse sequence symmetries are explored: \Rsymm431 (blue, left column) and \Rsymm873 (red, right column). The horizontal axis represents the rf field amplitude, expressed as a nutation frequency $\wnut$. The horizontal coordinates are given by the ratio $\wnut/\wnut^0$, where the nominal nutation frequency $\wnut^0$ is used to calculate the pulse durations, which are kept fixed. Row (a) shows that the \Rsymm431 and \Rsymm873 sequences are both fairly narrowband with respect to rf field amplitude when the standard \RNnnu protocol is used (figure~\ref{fig:Rseqstandard}). Row b shows that their robustness with respect to rf amplitude errors is greatly improved by the riffled variant of the \RNnnu protocol, inspired by PulsePol (figure~\ref{fig:RseqPP}). Their tolerance of rf amplitude errors is increased further when the central $180\deg$ pulses of the basic R-elements are replaced by ASBO-11 composite pulses~\cite{odedra_dual-compensated_2012} (row c). The use of $60_{180}180_{0}240_{180}420_{0}240_{180}180_{0}60_{180}$ composite pulses~\cite{shaka_symmetric_1987} provides less improvement (row d).  For comparison, the experimental performance of the M2S/S2M protocol~\cite{pileio_storage_2010,tayler_singlet_2011} is shown by the grey lines in row d. The performance of M2S/S2M is clearly inferior to that of the riffled \RNnnu sequences. 

Another important characteristic of pulse sequences for the generation and reconversion of singlet order is their robustness with respect to resonance offset, defined here as $\Delta\omega=\hf\omega_\Sigma$, where $\omega_\Sigma$ is the sum of the chemically shifted offset frequencies, see equation~\ref{eq:Hterms}. A robust performance with respect to resonance offset is usually desirable, since it renders the sequence less sensitive to inhomogeneity in the static magnetic field, which can be particularly important in low-field applications. 

Figure~\ref{fig:ResonanceOffsetDependence} compares the resonance-offset dependence of several pulse sequences, for the generation and reconversion of \Ctwo singlet order in the solution of \Ctwo-DAND. The left column compares different schemes which have \Rsymm431 symmetry. The right column compares different schemes which have \Rsymm873 symmetry. All experimental parameters are given in the Supporting Information. 

Figure~\ref{fig:ResonanceOffsetDependence}a shows the resonance-offset dependence of \RNnnu sequences constructed by the standard protocol of figure~\ref{fig:Rseqstandard}, using the basic R-element of equation~\ref{eq:R0}. The resulting sequences have a strong dependence on resonance offset, with the \Rsymm873 sequence displaying a particularly undesirable offset dependence. 

Figure~\ref{fig:ResonanceOffsetDependence}b shows the resonance-offset dependence of riffled \RNnnu sequences, using the pair of basic R-elements in equation~\ref{eq:R0AB}. Riffling clearly stabilises the resonance offset dependence, with the improvement being particularly striking for \Rsymm873. 

Figures~\ref{fig:ResonanceOffsetDependence}c and d explore the effect of substituting the central 180\deg pulse of the basic R-elements by composite pulses. Although
ASBO-11 composite pulses~\cite{odedra_dual-compensated_2012} do not change the performance of \Rsymm431 very much, they do lead to a significant increase in the bandwidth of \Rsymm873  (figure~\ref{fig:ResonanceOffsetDependence}c). An even more pronounced effect is observed upon replacing all single 180\deg pulses by 7-element $60_{180}180_{0}240_{180}420_{0}240_{180}180_{0}60_{180}$ composite pulses~\cite{shaka_symmetric_1987} (figure~\ref{fig:ResonanceOffsetDependence}d). The resonance-offset bandwidth of \Rsymm873 with 7-element composite pulses~\cite{shaka_symmetric_1987} is particularly impressive. 

The grey lines in figure~\ref{fig:ResonanceOffsetDependence}d show the experimental offset dependence of the M2S/S2M protocol~\cite{pileio_storage_2010}. All riffled \RNnnu sequences have a clearly superior performance to M2S/S2M. To put this in context, even the M2S/S2M protocol is regarded as relatively robust with respect to resonance offset, being first demonstrated on a sample in an inhomogeneous low magnetic field~\cite{pileio_storage_2010}. Some other techniques, such as SLIC~\cite{devience_preparation_2013}, are far more sensitive to resonance offset than M2S. 

Results for the dependence of the singlet order conversion on the pulse sequence intervals are given in the Supporting Information.

%===============
\section{\label{sec:Conclusions}Discussion}
%----------
The results shown in this paper indicate that PulsePol is a very attractive addition to the arsenal of pulse sequences for the manipulation of nuclear singlet order. The PulsePol sequences provide a high degree of robustness with respect to common experimental imperfections, which is found to be superior to existing methods such as M2S/S2M, especially when combined with composite pulses. This robustness is likely to be particularly important for applications to imaging and \emph{in vivo} experiments~\cite{berner_sambadena_2019,mamone_localized_2021}. 

In addition, PulsePol is a relatively simple repeating sequence of six pulses. This structure has many advantages over M2S, which performs the magnetization-to-singlet-order transformation in four consecutive steps~\cite{pileio_storage_2010,tayler_singlet_2011}. For example, the PulsePol repetitions may be stopped at any time, in order to achieve a partial transformation of spin order. This is more difficult to achieve for M2S and its variants. 

The theoretical relationship between PulsePol and symmetry-based recoupling sequences in solid-state NMR is unexpected. Nevertheless, this theoretical analogy immediately allows the considerable body of average Hamiltonian theory developed for symmetry-based recoupling to be deployed in this very different context. This immediately allows the use of symmetry-based selection rules for analysing existing PulsePol sequences and for designing new variants. 

All of the work reported in this paper uses the same set of basic elements, given in equations~\ref{eq:R0} and \ref{eq:R0AB}. There is clearly  scope for using different basic elements within the \RNnnu symmetry framework. 

As discussed above, PulsePol may interpreted as a variant implementation of \RNnnu symmetry, involving the alternation of two different basic elements, which compensate each others' imperfections. Such riffled \RNnnu sequences are more robust with respect to a range of experimental imperfections. The same principle might be applied to symmetry-based recoupling sequences in magic-angle-spinning solids. Extensions are also possible, involving more complex interleaved patterns of multiple basic elements. We intend to explore such ``riffled supercycles" in future work. 

In magic-angle-spinning solid-state NMR, symmetry-based pulse sequences have been used to address a wide variety of spin dynamical problems~\citeSBR, 
including multiple-channel sequences for the recoupling of heteronuclear systems~\cite{brinkmann_symmetry_2001,levitt_symmetry-based_2007}. Such extensions should be possible in the solution NMR context as well. 

Variants of M2S/S2M sequences have been applied to heteronuclear spin systems~\cite{eills_singlet_2017,stevanato_pulse_2017,bengs_robust_2020}. This has important applications in parahydrogen-induced polarization~\cite{eills_singlet_2017}. It is likely that riffled \RNnnu sequences are also applicable to this problem.

The theory of symmetry-based recoupling in magic-angle-spinning solids was originally formulated using average Hamiltonian theory, as sketched above. It is also possible to obtain the key results using Floquet theory~\citeFloquet, 
which may have advantages in certain circumstances. Floquet theory should also be applicable to the current context. 

In summary, the PulsePol sequence~\cite{schwartz_robust_2018,tratzmiller_pulsed_2021,tratzmiller_parallel_2021} is an important innovation that has potential applications in many forms of magnetic resonance. It sits at the fertile intersection of diamond magnetometry, quantum information processing, solid-state NMR, parahydrogen-induced hyperpolarization, and singlet NMR in solution. 

\begin{acknowledgments}
We acknowledge funding received by the European Research Council (grant 786707-FunMagResBeacons), and EPSRC-UK (grants EP/P009980/1, EP/P030491/1, EP/V055593/1). We thank Sami Jannin, Quentin Stern, Chlo{\'e} Gioiosa, Olivier Cala, Laurynas Dagys, and Maria Concistr{\'e} for help and discussions.
\end{acknowledgments}

\section*{Author Declarations}

\subsection*{Conflict of interest}

The authors have no conflicts to disclose.

\section*{Data Availability Statement}

The data that support the findings of this study are available from the corresponding author upon reasonable request.

\section*{References}
\bibliography{ms.bib}% 
%======================

\end{document}

\title{Symmetry-Based Singlet-Triplet Excitation in Solution Nuclear Magnetic Resonance}

\author{Mohamed Sabba}
\affiliation{Department of Chemistry, University of Southampton, SO17 1BJ, UK}

\author{Nino Wili}
\affiliation{Interdisciplinary Nanoscience Center (iNANO) and Department of Chemistry, Aarhus University, Gustav Wieds Vej 14, DK-8000 Aarhus C, Denmark}

\author{Christian Bengs}
\affiliation{Department of Chemistry, University of Southampton, SO17 1BJ, UK}

\author{Lynda J. Brown}
\affiliation{Department of Chemistry, University of Southampton, SO17 1BJ, UK}

\author{Malcolm H. Levitt}
 \email{mhl@soton.ac.uk}
 \affiliation{Department of Chemistry, University of Southampton, SO17 1BJ, UK}

\date{\today}

\maketitle

%%%%%%%
%======
%%%%%%%

%================================================
\section{Pulse Sequence Details}

\subsection{Composite pulses}

\subsubsection{BB1 composite pulse}
The BB1 family of composite pulses originally defined by Wimperis~\cite{wimperis_broadband_1994} achieves broadband compensation of pulse strength errors. In the time-symmetric version~\cite{cummins_tackling_2003}, which we designate $BB1(\beta)$, a composite implementation of a simple $\beta_0$ pulse with generic flip angle $\beta$ takes the following form:
\begin{equation}
BB1(\beta) = (\beta/2)_{0}180_{\theta_{W}(\beta)}360_{3\theta_{W}(\beta)} 180_{\theta_{W}(\beta)}(\beta/2)_{0}
\end{equation}

The angle $\theta_{W}$ in the phases of the error correcting block depends on the desired flip angle $\beta$, and is given by:
\begin{equation}
\theta_{W}(\beta) =\arccos({-\beta /(4\pi))} = \arccos({-\beta /(720^{\circ}))}
\end{equation}
\\
For a $90^{\circ}$  and $180^{\circ}$ pulse respectively: 
\begin{equation}
\theta_{W}(\pi/2) = \arccos{(-1/8)} \approx 97.18^{\circ}
\end{equation}
\begin{equation}
\theta_{W}(\pi) = \arccos{(-1/4)} \approx 104.48^{\circ}
\end{equation}

Accordingly, in all our singlet-filtered experiments, the $90_{0}$ readout pulses at the end are replaced by the equivalent composite  rotation $45_{0}180_{97.18}360_{291.54}180_{97.18}45_{0}$. Additionally, a two step [0,180] phase cycle is implemented on the readout pulse and receiver channel.

\subsubsection{ASBO-11 composite pulse}
ASBO-11 is a closely related infinite family of \emph{dual-compensated} composite inversion pulses which achieves simultaneous compensation of pulse strength errors and resonance offset/detuning errors. It replaces a single $180_{0}$ pulse with 11 180 pulses with phases arranged in a so-called \emph{antisymmetric} (i.e. the time reverse inverts all phases) form such as:

\begin{equation}
\rm{\pi^{11}_{ASBO}} = 180_{-\phi_1}180_{-\phi_2}180_{-\phi_3}180_{-\phi_4}180_{-\phi_5}180_{0}180_{+\phi_5}180_{+\phi_4}180_{+\phi_3}180_{+\phi_2}180_{+\phi_1}
\end{equation}

In general, the phases of the 11 pulses are given by:

\begin{equation}
\phi_{1}=\tfrac{2}{3}\pi -5\phi
\end{equation}
\begin{equation}
\phi_{2}=\tfrac{4}{3}\pi - \theta_{W}(\pi) -4\phi
\end{equation}
\begin{equation}
\phi_{3}=\tfrac{4}{3}\pi -
2\theta_{W}(\pi)- 3\phi
\end{equation}
\begin{equation}
 \phi_{4}=\tfrac{4}{3}\pi - \theta_{W}(\pi) - 2\phi  
\end{equation}
\begin{equation}
\phi_{5}=\tfrac{2}{3}\pi -\phi
\end{equation}

In this context, $\phi$ is a free variable which may be tailored for the compensation of resonance offset errors, pulse strength errors, or both.

We have found that the choice $\phi = \tfrac{4}{3}\pi -\theta_{W}(\pi)/2 \approx 187.8^{\circ}$ works well for dual-compensation. This choice of phase appears to correspond to "ASBO-11($B_{1}$)" described by Odedra et al. (they give $\phi = 188^{\circ}$) which was found by a numerical search over $\phi$ in $1^{\circ}$ increments for the ASBO-11 sequence with the largest bandwidth with respect to pulse strength errors.
\\
For $\phi = \tfrac{4}{3}\pi - \theta_{W}(\pi)/2$, we obtain the set of solutions:

\begin{equation}
    ({\phi_1},{\phi_2},{\phi_3},{\phi_4},{\phi_5}) = ({\frac{5}{2}\theta_{W}(\pi),\theta_{W}(\pi),\tfrac{4}{3}\pi -\theta_{W}(\pi)/2,\tfrac{2}{3}\pi ,\tfrac{4}{3}\pi  +\theta_{W}(\pi)/2})
\end{equation}
\\
Accordingly, this leads to the ASBO-11 composite pulse tested in our experiments:
\begin{equation}
180_{98.81}180_{255.52}180_{172.24}180_{240}180_{67.76}180_{0}180_{292.24}180_{120}180_{187.76}180_{104.45}180_{261.19}    
\end{equation}

\subsection{$T_{00}$ filter}

The $T_{00}$ filter is a common block in singlet NMR experiments. It consists of a series of pulsed field gradients and radiofrequency pulses which are designed to dephase unwanted operators i.e. those not corresponding to the $T_{00}$ symmetry of the nuclear singlet order operator. A typical implementation consists of three gradients sandwiched by two radiofrequency pulses:
\begin{equation}
    G_{1}-90_0-G_{2}-{\beta_{m}}_{0}-G_{3}
\end{equation}
Here, the angle ${\beta_{m}}$ is the magic angle $\arctan{\sqrt{2}} \approx 54.74\deg$. 
\begin{figure}[tbh]
\caption{Illustration of the $T_{00}$ filter implemented in experiments in the main text.}
\centering
\includegraphics[scale=0.3]{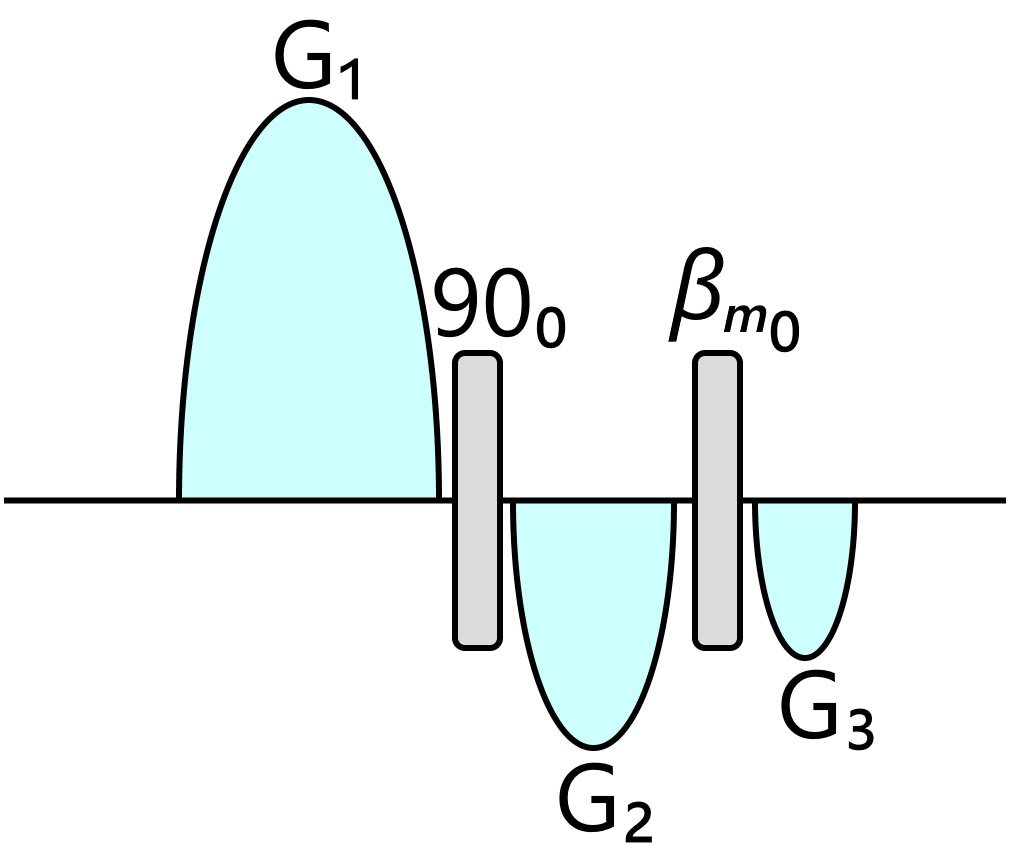}
\end{figure}
In order to ensure the optimal performance of the $T_{00}$ filter, all pulses were replaced by the corresponding BB1 composite pulses as described in the previous subsection; the 90\deg pulse is implemented as $45_{0}180_{97.18}360_{291.54}180_{97.18}45_{0}$ while the  $(\beta_{m})_{0}$ pulse is implemented as $27.37_{0}180_{94.36}360_{283.08}180_{94.36}27.37_{0}$. 
\\
\\
The parameters used in our experiments are shown in Table SI. In practice, due to hardware limitations, rest delays $\tau_r$ follow each pulsed field gradient.
 \begin{table}[!htb]
  \label{tab:T_zz params}
\caption{Experimental parameters for the $T_{00}$ filter used in the experiments. The gradient strengths are given by $G_1$, $G_2$, and $G_3$ respectively. The gradient durations are given by $\tau^{G}_{1}$, $\tau^{G}_{2}$, and $\tau^{G}_{3}$ respectively. The recovery delay after each gradient is given by $\tau^{rest}_{1}$, $\tau^{rest}_{2}$, and $\tau^{rest}_{3}$.
  }
  \begin{tabular*}{0.4\textwidth}{@{\extracolsep{\fill}}lll}
    \hline
    $G_{1} [\rm{G/cm}]$ & 16.08
    \\
    $G_{2} [\rm{G/cm}]$ &  -9.94
    \\
    $G_{3} [\rm{G/cm}]$ & -6.14
    \\
    $\tau^{G}_{1} \rm{[\mu s]}$ &  8000.000
    \\
    $\tau^{G}_{2} \rm{[\mu s]}$ & 4944.272
    \\
    $\tau^{G}_{3} \rm{[\mu s]}$ & 3055.728
    \\
    $\tau^{rest}_{1} [\rm{ms}]$ &  20.4
    \\
    $\tau^{rest}_{2} [\rm{ms}]$ & 15.4
    \\
    $\tau^{rest}_{3} [\rm{ms}]$ & 17.3
    \\
    \hline
  \end{tabular*}
\end{table}

\subsection{Singlet order destruction (SOD) element}
In standard NMR experiments, the waiting delay between scans is typically set to be on the order of $\times5$ the longitudinal relaxation constant $T_1$, which is usually enough to fully equilibrate a spin system for most practical purposes. However, in experiments which excite nuclear singlet order - which relaxes with a time constant $T_S$, often orders of magnitude larger than $T_1$ - this approach is problematic. 
\\
In order to ensure the quality of experimental data, a singlet order destruction (SOD) element was incorporated in all experiments.

The SOD element consists of a $T_{00}$ filter followed by a train of J-synchronized spin echoes repeated $m_1$ times.
\\
The J-synchronized block is a building block of M2S, and similar to the M2S sequence has a total echo duration $\tau_{e}$ ideally set to:
\begin{equation}
    \tau_{e} = 1/(2J)
\end{equation}
For optimal singlet order destruction, the number of repetitions should roughly accomplish a $2\pi/3$ rotation in the $\ket{S_0}$-$\ket{T_0}$ Bloch sphere~\cite{rodin_SOD_2019}:
\begin{equation}
    m_{1} \approx round({\pi/(3\theta_{ST})})
\end{equation}
The SOD element may be repeated $m_2$ times. Previous work\cite{rodin_SOD_2019} suggests $m_2 \approx 1-3$ is sufficient for singlet order destruction. Out of an abundance of caution, we set $m_2 = 7$ in our experiments.

The SOD element is illustrated in Figure S2.
\begin{figure}[!htb]
\caption{Illustration of the SOD filter implemented in the experiments. The $T_{00}$ filter has the same meaning as the previous section. $\tau_e$ is the total spin echo duration. $m_1$ is the number of times the spin echo is repeated within a single SOD element. $m_2$ is the total number of SOD elements. $\tau_r$ is the relaxation delay.}
\centering
\includegraphics[scale=0.3]{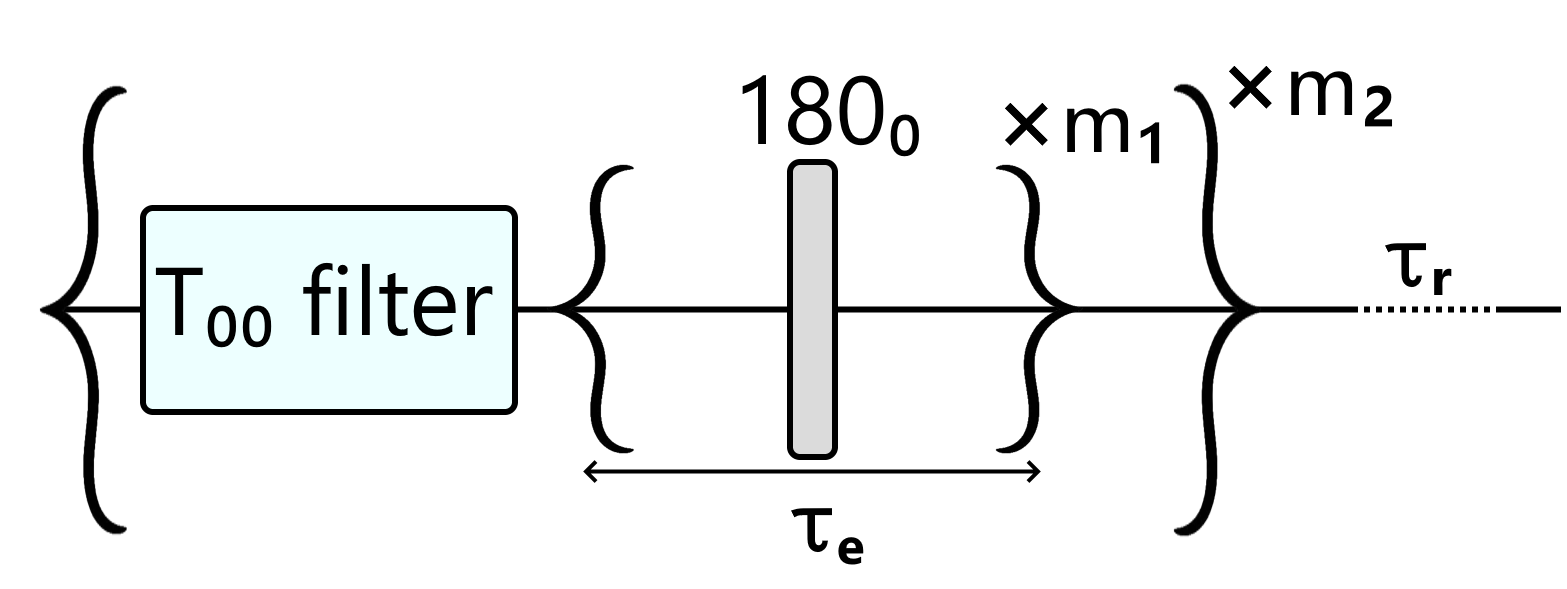}
\end{figure}
\\
The parameters used in the SOD element in the main text are given in Table SII.

 \begin{table}[tbh]
  \label{tab:T_zz params}
\caption{Experimental parameters for the SOD block used in the experiments in the main text. the parameters have the same meaning as Figure 2.
  }
  \begin{tabular*}{0.4\textwidth}{@{\extracolsep{\fill}}lll}
    \hline
    $m_{1}$ & 7
    \\
    $m_{2}$ & 7
    \\
    $\tau_{e} [\rm{ms}]$ & 9.24
    \\
    $\tau_{r} [\rm{s}]$ &  30
    \\
    \hline
  \end{tabular*}
\end{table}
\pagebreak
\section{Experimental details for figures 6-9}
\subsection{Description of M2S/S2M sequences}

The M2S sequence is prototypical hard-pulse sequence for generating singlet order from longitudinal magnetization in the near-equivalence regime~\cite{pileio_storage_2010,tayler_singlet_2011,tayler_theory_2012}. In general, M2S takes the form:
\begin{equation}
    90_{x}-(\tau_{1}-90_{y}180_{x}90_{y}-\tau_{1})^{n_1}-90_{y}-\tau_{2}-(\tau_{1}-90_{y}180_{x}90_{y}-\tau_{1})^{n_2}
\end{equation}
Here, $\tau_1$ and $\tau_2$ are interpulse delays, while $n_1$ and $n_2$ denote the number of repetitions.
\begin{figure}[!htb]
\caption{Illustration of the M2S sequence in this work. $\tau_1$ is the interval between pulses in the spin echoes (of total duration $\tau_e$), and $\tau_2$ is the interval after the $90_{y}$ pulse.  
}
\centering
\includegraphics[scale=0.3]{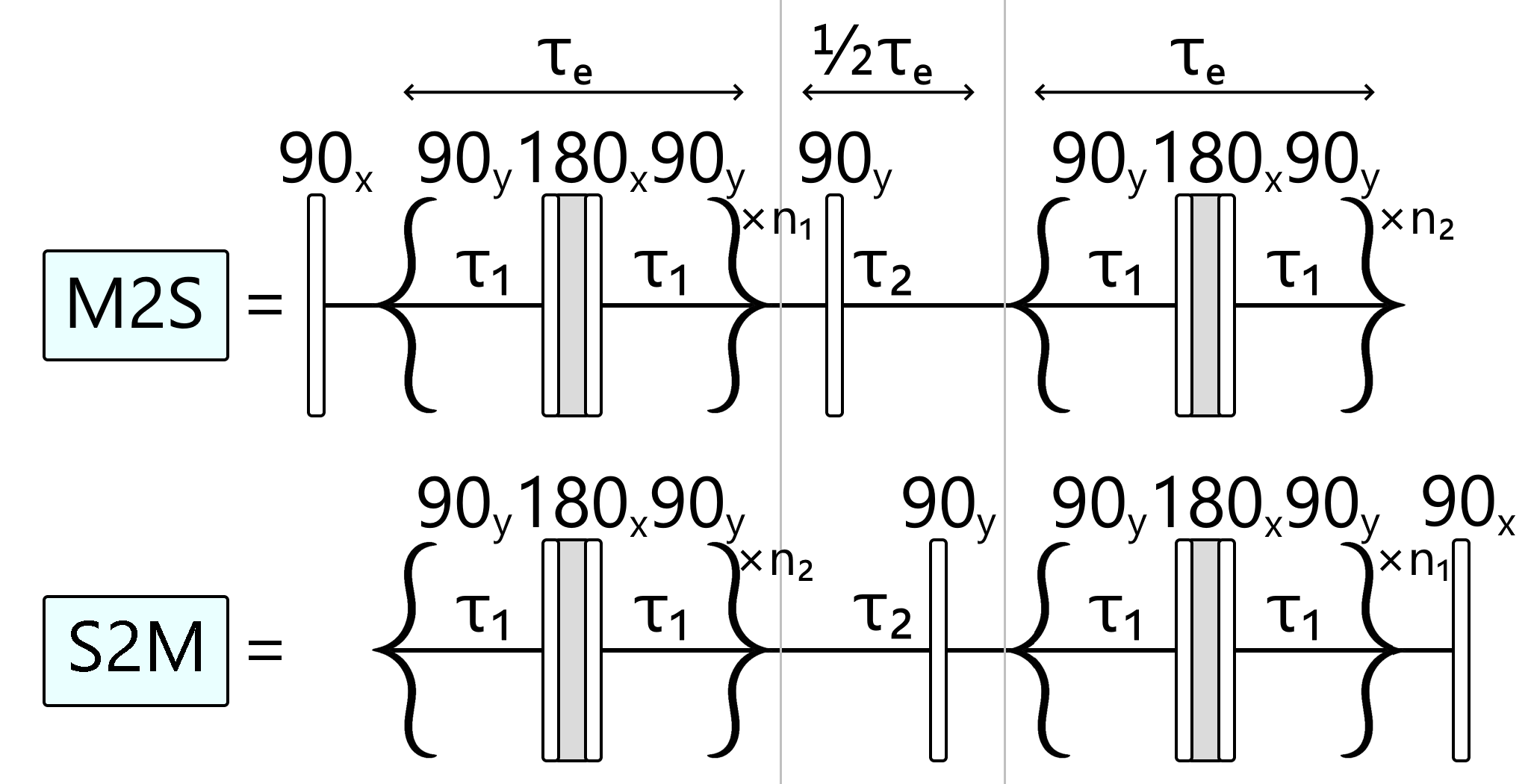}
\end{figure}
\\
Unlike the simple presentation of an R-sequence, M2S consists of five distinct blocks: 
(i) a 90\deg excitation pulse; (ii) a train of $n_{1}$ J-modulated spin echoes of total duration $\tau_{e} \approx 1/(2J)$; 
(iii) another 90\deg pulse with a phase in quadrature with the initial excitation pulse; 
(iv) a $\tau_{e}/2 \approx 1/(4J)$ refocusing delay;
(v) a train of $n_{2}\approx n_{1}/2$ J-modulated spin echoes.

The pulse sequence which reconverts singlet order to magnetization is the emph{time reverse}, denoted S2M.

To ensure maximum error compensation, the 180\deg pulses in the echo trains are implemented with the standard MLEV-4 four-step [0,0,180,180] supercycle~\cite{levitt_broadband_1982,pileio_storage_2010,tayler_theory_2012}.

\subsection{Parameters for sequences in Figure 6}

The experimental parameters for the \RNnnu and M2S sequences that appear in Figure 6 are shown in Table SIII.

 \begin{table}[htb!]
 \renewcommand{\arraystretch}{0.6}
  \begin{tabular*}{0.6\textwidth}{@{\extracolsep{\fill}}lll}
\hline
    $\omega_{nut}/(2\pi)$ & $12.5$ kHz
\\
    $\tau_{90}$ & 20 $\mu$s
\\
\hline
\multirow{4}{4em}{\Rsymm431 (riffled)}
 &   $\tau_R$    & 13800 ${\,\mu \mathrm{s}}$
\\
 &   $\tau$    & 6860 ${\,\mu \mathrm{s}}$
\\
 &   $n_R^\mathrm{exc}$       &  9
\\
 &   $\tau_\mathrm{exc}$     & 124.2 ms
 \\ \hline
 \multirow{4}{4em}{\Rsymm431 (standard)}
 &   $\tau_R$    & 13400 ${\,\mu \mathrm{s}}$
\\
 &   $\tau$    & 6660 ${\,\mu \mathrm{s}}$
\\
 &   $n_R^\mathrm{exc}$       &  9
\\
 &   $\tau_\mathrm{exc}$     & 120.60 ms
\\ \hline
\multirow{4}{4em}{\Rsymm873 (riffled)} 
 &   $\tau_R$    & 16000 ${\,\mu \mathrm{s}}$
\\
 &   $\tau$    & 7960 ${\,\mu \mathrm{s}}$
\\
 &   $n_R^\mathrm{exc}$       &  9
\\
 &   $\tau_\mathrm{exc}$     & 144.00 ms
  \\ \hline
 \multirow{4}{4em}{\Rsymm873 (standard)}
 &   $\tau_R$    & 15560 ${\,\mu \mathrm{s}}$
\\
 &   $\tau$    & 7740 ${\,\mu \mathrm{s}}$
\\
 &   $n_R^\mathrm{exc}$       &  9
\\
 &   $\tau_\mathrm{exc}$     & 140.04 ms
 \\
\hline
\multirow{4}{4em}{M2S} 
 &   $\tau_\mathrm{e}$    & 9240 ${\,\mu \mathrm{s}}$
\\
 &   $\tau_1$    & 4580 ${\,\mu \mathrm{s}}$
 \\
 &   $\tau_2$    & 4600 ${\,\mu \mathrm{s}}$
\\
 &   $n_1$       &  11
\\
 &   $n_2$       & 5
 \\
 &   $\tau_\mathrm{exc}$     & 152.46 ms
\\ 
\hline
  \end{tabular*}
 \caption{
 \label{tab:RExcPars}
 Experimental parameters for the M2S and \RNnnu sequences used to obtain the results in Figure 6(b,c,d,e,f) in the main text. The parameters for the \RNnnu sequences have
 the same meaning as in Table III in the main text. The parameters are given separately for \Rsymm431 sequences (used in Figure 6(c,e)), the \Rsymm873 sequences (used in Figure~6(d,f)), and the M2S sequence (used in Figure 6(b).)}
 \end{table}

\subsection{Parameters for sequences in Figure 7}

 \begin{table}[htb!]
 \renewcommand{\arraystretch}{0.6}
  \begin{tabular*}{0.6\textwidth}{@{\extracolsep{\fill}}lll}
\hline
    $\omega_{nut}/(2\pi)$ & $12.5$ kHz
\\
    $\tau_{90}$ & 20 $\mu$s
\\
\hline
\multirow{2}{8em}{\Rsymm431 (riffled)}
 &   $\tau_R$    & 13800 ${\,\mu \mathrm{s}}$
\\
 &   $\tau$    & 6860 ${\,\mu \mathrm{s}}$
 \\ \hline
\multirow{2}{8em}{\Rsymm873 (riffled)} 
 &   $\tau_R$    & 16000 ${\,\mu \mathrm{s}}$
\\
 &   $\tau$    & 7960 ${\,\mu \mathrm{s}}$
  \\ \hline
\multirow{2}{8em}{\Rsymm{10}32 (riffled)} 
 &   $\tau_R$    & 5560 ${\,\mu \mathrm{s}}$
\\
 &   $\tau$    & 2720 ${\,\mu \mathrm{s}}$
\\ \hline
  \end{tabular*}
 \caption{
 \label{tab:RExcPars}
 Experimental parameters for the \RNnnu sequences used to obtain the results in Figure 7 in the main text. The parameters for the \RNnnu sequences have
 the same meaning as in Table SIII. The parameters are given separately for the \Rsymm431 sequence (used in Figure 7(a), the \Rsymm873 sequence (used in Figure~7(b)), and the \Rsymm{10}32 sequence (used in Figure~7(c)).}
 \end{table}
\newpage
\subsection{Parameters for sequences in Figures 8 and 9}
 \begin{table}[htb!]
 \renewcommand{\arraystretch}{0.6}
  \begin{tabular*}{0.5\textwidth}{@{\extracolsep{\fill}}lll}
\hline
    $\omega_{nut}^{0}/(2\pi)$ & $12.5$ kHz
\\
    $\tau_{90}$ & 20 $\mu$s
\\
\hline
 \multirow{4}{8em}{\Rsymm431 (standard)}
 &   $\tau_R$    & 13400 ${\,\mu \mathrm{s}}$
\\
 &   $\tau$    & 6660 ${\,\mu \mathrm{s}}$
\\
 &   $n_R^\mathrm{exc}$       &  9
\\
 &   $\tau_\mathrm{exc}$     & 120.60 ms
\\ \hline
 \multirow{4}{8em}{\Rsymm431 (riffled)}
 &   $\tau_R$    & 13800 ${\,\mu \mathrm{s}}$
\\
 &   $\tau$    & 6860 ${\,\mu \mathrm{s}}$
\\
 &   $n_R^\mathrm{exc}$       &  9
\\
 &   $\tau_\mathrm{exc}$     & 124.2 ms
 \\ \hline
 \multirow{4}{8em}{\Rsymm431 (ASBO-11)}
 &   $\tau_R$    & 13800 ${\,\mu \mathrm{s}}$
\\
 &   $\tau$    & 6460 ${\,\mu \mathrm{s}}$
\\
 &   $n_R^\mathrm{exc}$       &  9
\\
 &   $\tau_\mathrm{exc}$     & 124.2 ms
 \\ \hline
  \multirow{4}{8em}{\Rsymm431 (SP7)}
 &   $\tau_R$    & 13800 ${\,\mu \mathrm{s}}$
\\
 &   $\tau$    & 6593 ${\,\mu \mathrm{s}}$
\\
 &   $n_R^\mathrm{exc}$       &  9
\\
 &   $\tau_\mathrm{exc}$     & 124.2 ms
 \\ \hline
 \multirow{4}{8em}{\Rsymm873 (standard)}
 &   $\tau_R$    & 15560 ${\,\mu \mathrm{s}}$
\\
 &   $\tau$    & 7740 ${\,\mu \mathrm{s}}$
\\
 &   $n_R^\mathrm{exc}$       &  9
\\
 &   $\tau_\mathrm{exc}$     & 140.04 ms
 \\ \hline
 \multirow{4}{8em}{\Rsymm873 (riffled)} 
 &   $\tau_R$    & 16000 ${\,\mu \mathrm{s}}$
\\
 &   $\tau$    & 7960 ${\,\mu \mathrm{s}}$
\\
 &   $n_R^\mathrm{exc}$       &  9
\\
 &   $\tau_\mathrm{exc}$     & 144.00 ms
  \\ \hline
 \multirow{4}{8em}{\Rsymm873 (ASBO-11)}
 &   $\tau_R$    & 16000 ${\,\mu \mathrm{s}}$
\\
 &   $\tau$    & 7560 ${\,\mu \mathrm{s}}$
\\
 &   $n_R^\mathrm{exc}$       &  9
\\
 &   $\tau_\mathrm{exc}$     & 144.00 ms
  \\ \hline
   \multirow{4}{8em}{\Rsymm873 (SP7)}
 &   $\tau_R$    & 16000 ${\,\mu \mathrm{s}}$
\\
 &   $\tau$    & 7693 ${\,\mu \mathrm{s}}$
\\
 &   $n_R^\mathrm{exc}$       &  9
\\
 &   $\tau_\mathrm{exc}$     & 144.00 ms
  \\ \hline
  \end{tabular*}
 \caption{
 \label{tab:RExcPars}
 Experimental parameters for the \RNnnu sequences used to obtain the results in Figures 8(a,b,c,d) and 9(a,b,c,d) in the main text. The parameters for the \RNnnu sequences have
 the same meaning as in Tables SIII-IV. The parameters are given separately for \Rsymm431 and \Rsymm873 sequences in the standard implementation (Figures 8(a) and 9(a)); the riffled implementation (Figures 8(b) and 9(b)); the riffled implementation with the ASBO-11 composite pulse (Figures 8(c) and 9(c)); and the riffled implementation with the 7-element Shaka-Pines\cite{shaka_symmetric_1987} (SP7) composite pulse (Figures 8(d) and 9(d)).}
 \end{table}

\section{Relaxation Experiments}
\subsection{$T_{1}$ measurement}
The time constant for the relaxation of longitudinal magnetization is typically denoted $T_1$ in NMR.

We have used a standard inversion recovery experiment to measure T1, as shown in Figure S4.

\begin{figure}[!htb]
\caption{Illustration of the inversion recovery sequence used to measure $T_{1}$. After a relaxation delay of 30 seconds, the longitudinal magnetization is inverted with a composite pulse, allowed to evolve, and then read out with a 90 degree pulse.}
\centering
\includegraphics[scale=0.3]{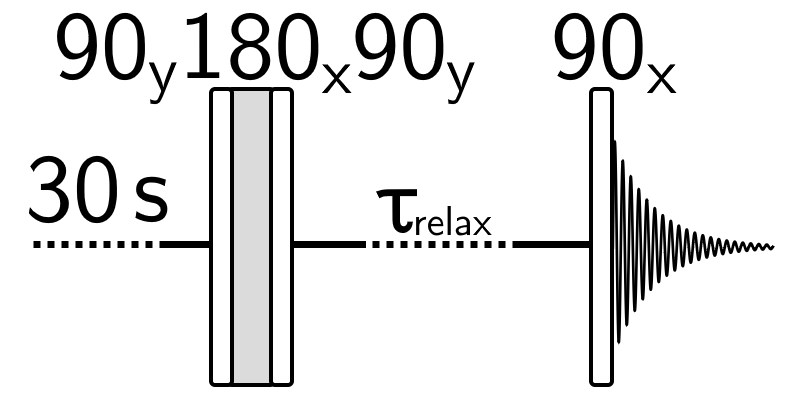}
\end{figure}

The time evolution of magnetization following inversion, $M(t)$, may be fitted to the simple equation:

\begin{equation}
    M(t) = A(1-2\exp{(-t/T_{1})})
\end{equation}

\begin{figure}[!htb]
\caption{Longitudinal relaxation of spin magnetization in $^{13}C_{2}$-DAND@ 9.4 T and 25 \deg C, following the experiment in Figure 3.. Black circles: experimental data. Dashed line: fit using Equation (16), with the parameters $A = 0.98 4\pm 0.006$ and $T_ 1 = 3.41 \pm 0.05 \rm{ s} $}
\centering
\includegraphics[scale=0.5]{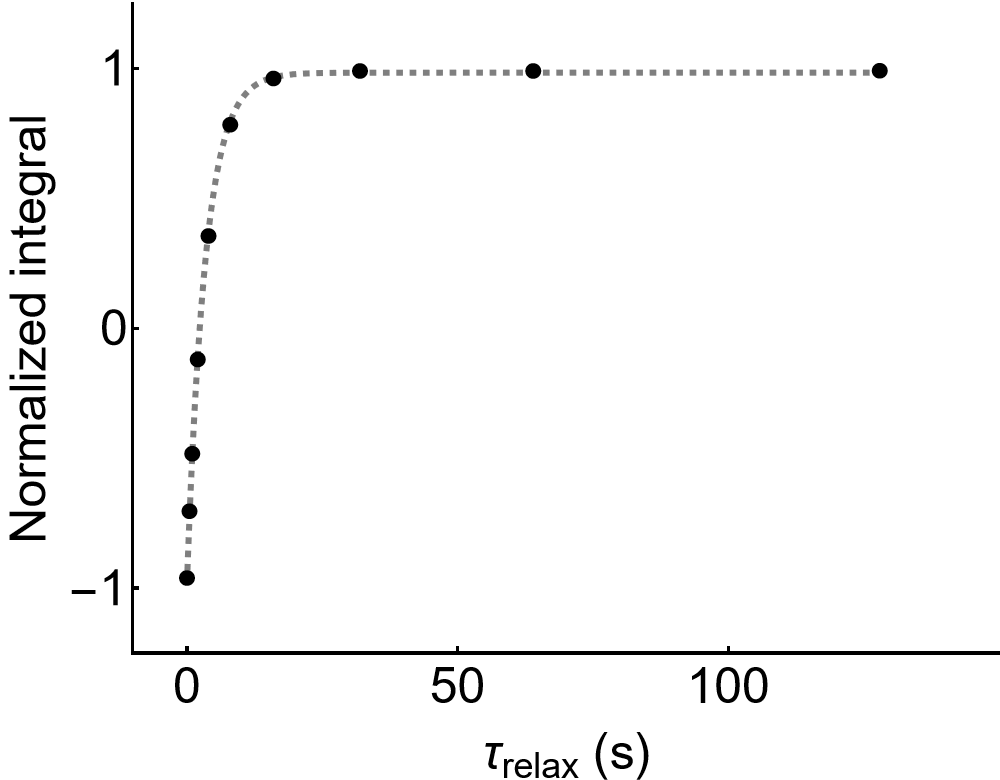}
\end{figure}

\subsection{$T_{S}$ measurement using PulsePol}
The singlet relaxation time $T_{S}$ can be measured using the sequences described in the main text.

\begin{figure}[!htb]
\caption{Illustration of the inversion recovery sequence used to measure $T_{S}$. After the SOD filter, and generation of nuclear singlet order using the $R4^{1}_{3}$ sequence, the singlet order is allowed to evolve, filtered, and then read out with another $R4^{1}_{3}$ sequence and a 90 degree pulse. The $R4^{1}_{3}$ sequence is performed as per the PulsePol implementation, and has the parameters described in the main text.}
\centering
\includegraphics[scale=0.3]{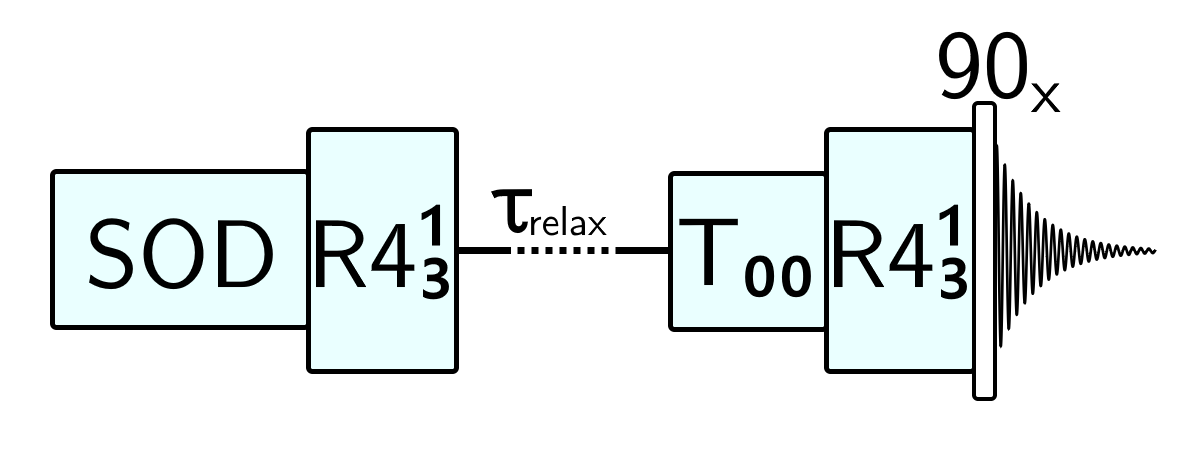}
\end{figure}

The time evolution of nuclear singlet order may be fitted to the simple equation:

\begin{equation}
    M(t) = A\exp{(-t/T_{S})}
\end{equation}

\begin{figure}[!htb]
\caption{Singlet relaxation in TEMPO-doped $^{13}C_{2}$-DAND solution @ 9.4 T and 25\deg C following the experiment in Figure 5. Black circles: experimental data. Dashed line: fit using Equation (17), with the parameters $A = 1.03 \pm 0.01$ and $T_ S = 89.4 \pm 4.3 \rm{ s} $}
\centering
\includegraphics[scale=0.5]{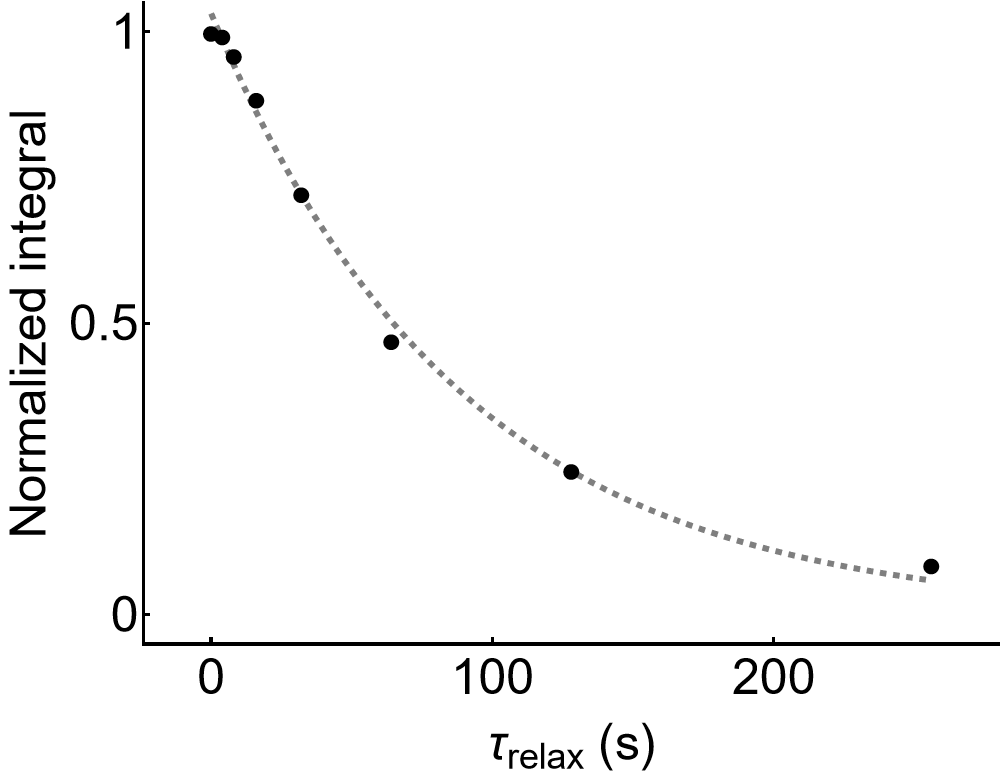}
\end{figure}
\clearpage
\section{Additional Performance Comparisons}
\subsection{Dependence on delay mismatch}
\begin{figure}[htb!]
\caption{\label{fig:tauvar} Experimental $^{13}{\rm C}$ signal amplitudes (white dots) for (a) \Rsymm431, (b) \Rsymm873 and (c) M2S as a function of the relative inter-pulse delay mismatch $\Delta \tau/\tau^{0}$, where $\tau^{0}$ represents the nominal inter-pulse delay. For the M2S sequence the nominal inter-pulse delay is given by $\tau^{0}=1/(4 J)$, whereas for R-based sequences the nominal inter-pulse delay is given by $\tau^{0}=n/(N J)$. The R-sequences have been implemented according to the PulsePol procedure. The final $^{13}{\rm C}$ signal amplitudes were referenced with respect to a single $^{13}{\rm C}$-pulse-acquire spectrum. Light blue trajectories represent numerical simulations with the pulse sequence parameters given in Tables I-II. Relaxation was neglected in all cases.
}
\centering
\includegraphics[width=0.5\columnwidth]{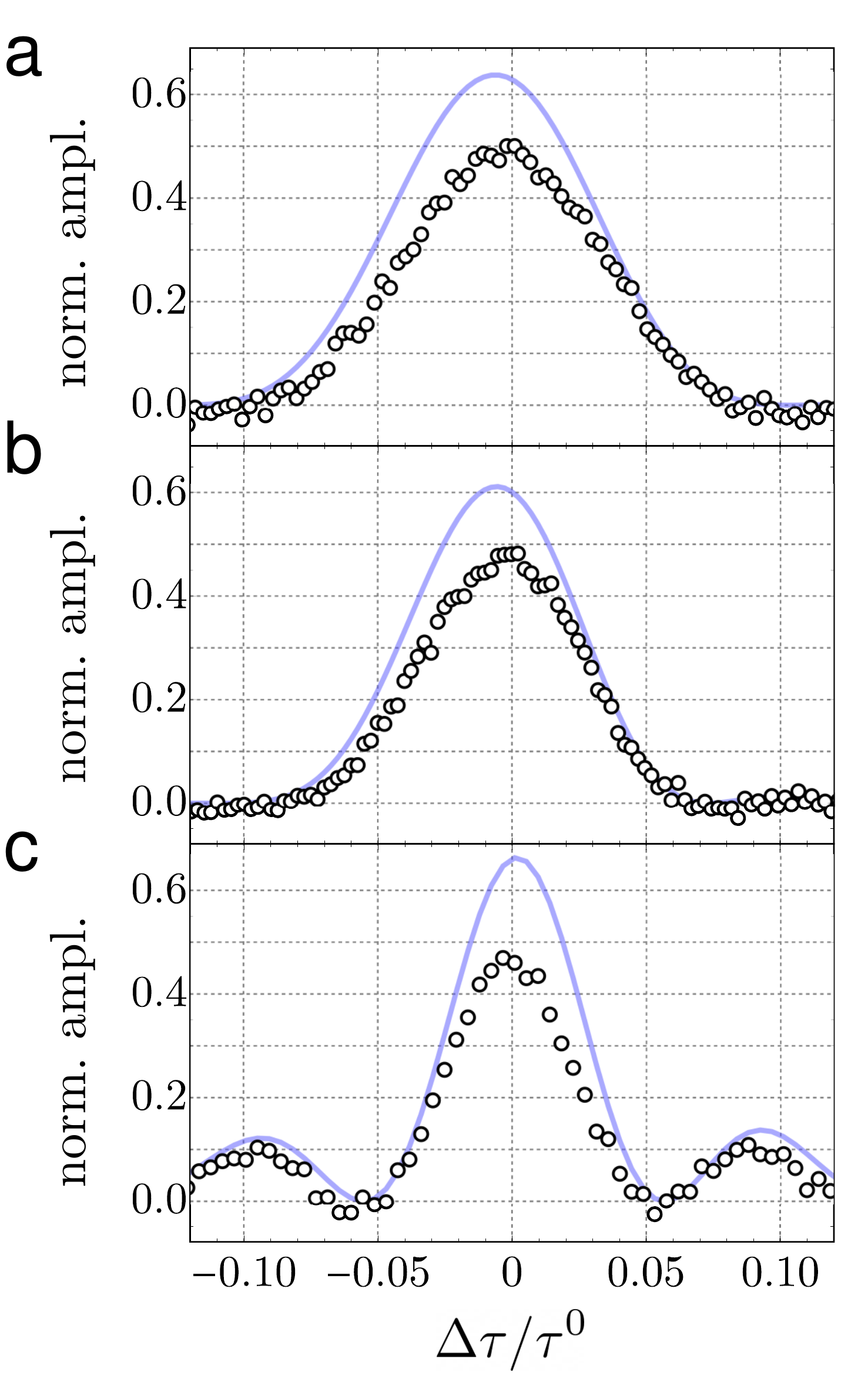}
\end{figure} 

\
\
\section{references}
\bibliography{supplement.bib}%